\documentclass[twocolumn]{aastex701}
\usepackage{siunitx} 

\newcommand{\p}{\partial}

\newcommand{\kms}{\ensuremath{\mathrm{km\,s^{-1}}}}
\newcommand{\cmc}{\ensuremath{\mathrm{cm^{-3}}}}
\newcommand{\cms}{\ensuremath{\mathrm{cm^{-2}}}}
\newcommand{\K}{\ensuremath{\mathrm{K}}}
\newcommand{\mK}{\ensuremath{\mathrm{mK}}}
\newcommand{\pc}{\ensuremath{\mathrm{pc}}}

\newcommand{\GHz}{\ensuremath{\mathrm{GHz}}}
\newcommand{\persec}{\ensuremath{\mathrm{s^{-1}}}}
\newcommand{\MHz}{\ensuremath{\mathrm{MHz}}}
\newcommand{\kmspc}{\ensuremath{\mathrm{km\,s^{-1}\,pc^{-1}}}}

\begin{document}

\title{What Heats the Dense Gas in the Galactic Center?}

\author[0009-0009-3431-1150,gname=Zhenyi,sname=Yue]{Zhenyi Yue}
\affiliation{Purple Mountain Observatory, Chinese Academy of Science, 10 Yuanhua Road, Nanjing 210023, People’s Republic of China}
\affiliation{School of Astronomy and Space Sciences, University of Science and Technology of China, Hefei 230026, People’s Republic of China}
\email{zyyue@pmo.ac.cn}

\author[0000-0003-3139-2724,gname=Yiping, sname=Ao]{Yiping Ao}
\affiliation{Purple Mountain Observatory, Chinese Academy of Science, 10 Yuanhua Road, Nanjing 210023, People’s Republic of China}
\affiliation{School of Astronomy and Space Sciences, University of Science and Technology of China, Hefei 230026, People’s Republic of China}
\email[show]{ypao@pmo.ac.cn}

\author[0000-0002-4154-4309,gname=Xindi, sname=Tang]{Xindi Tang}
\affiliation{Xinjiang Astronomical Observatory, Chinese Academy of Sciences, Urumqi 830011, People’s Republic of China}
\email{tangxindi@xao.ac.cn}

\author[0000-0003-2619-9305,gname=Xing, sname=Lu]{Xing Lu }
\affiliation{Shanghai Astronomical Observatory, Chinese Academy of Sciences, 80 Nandan Road, Shanghai 200030, People’s Republic of China}
\email{xinglu@shao.ac.cn}

\author[0000-0002-3866-414X,gname=Yan, sname=Gong]{Yan Gong}
\affiliation{Purple Mountain Observatory, Chinese Academy of Science, 10 Yuanhua Road, Nanjing 210023, People’s Republic of China}
\affiliation{School of Astronomy and Space Sciences, University of Science and Technology of China, Hefei 230026, People’s Republic of China}
\email{ygong@pmo.ac.cn}

\author[0000-0002-7495-4005, gname=Christian, sname=Henkel]{Christian Henkel}
\affiliation{Max-Planck-Institut für Radioastronomie, Auf dem Hügel 69, 53121 Bonn, Germany}
\affiliation{Xinjiang Astronomical Observatory, Chinese Academy of Sciences, Urumqi 830011, People’s Republic of China}
\email{chenkel@mpifr-bonn.mpg.de}

\begin{abstract}
Previous studies using p-H$_2$CO $J=3$--$2$ transitions at 218~\rm{GHz} suggested widespread high-temperature gas exceeding 60~\K{} and even 100~\K{} in the CMZ, with heating mechanisms possibly related to cosmic rays or turbulent dissipation. However, at temperatures above 100~\K{}, p-H$_2$CO $J=3$--$2$ line emission may lead to significant overestimates of kinetic temperature. This study combines o-H$_2$CO $J=5$--$4$ data from JCMT with p-H$_2$CO $J=3$--$2$ data from APEX to analyze three molecular clouds (The Brick, Sgr~A1, and Sgr~A2) with high temperatures. We used the non-LTE radiative transfer code RADEX to model spectral lines and constrain physical parameters with multiple line ratios, obtaining more reliable kinetic temperatures. Our results show that the previously reported extreme temperatures ($>100$~\K{}) based on p-H$_2$CO $J=3$--$2$ line ratios are revised downward, with the average kinetic temperatures now constrained to 84--95~\K{} using o-H$_2$CO $J=5$--$4$ line ratios, indicating systematic overestimation in the earlier studies. Further analysis reveals that the relationship between temperature and gas line width aligns more closely with predictions from models incorporating both high cosmic ray ionization rate and turbulent heating, suggesting that these molecular clouds are likely heated by a combination of cosmic-ray and turbulent dissipation mechanisms.

\end{abstract}


\section{Introduction}
\label{sec:introduction}
\setcounter{footnote}{0}
The Central Molecular Zone (CMZ) of our Galaxy hosts a unique and extreme interstellar environment \citep[e.g.,][]{Morris1996}, characterized by physical conditions that differ markedly from those of giant molecular clouds in the Galactic disk. These include extremely high gas densities ($n$(H$_2$) $\sim 10^3$--$10^6$~\cmc{}; \citealt{Longmore2013, Mills2018, Battersby2025}), high Mach numbers ($M_s > 10$; \citealt{Henshaw2016, Federrath2016}), elevated gas and dust temperatures ($T_{\mathrm{gas}} \sim 50$--$100$~\K{}, $T_{\mathrm{dust}} \sim 20$--$50$~\K{}; \citealt{Morris1983, Guesten1985, Ao2013, Mills2013, Ott2014, Ginsburg2016}), strong magnetic fields ($B \lesssim 1$~$\mathrm{mG}$; \citealt{Spergel1992, Lu2024, Pare2024, Butterfield2024, Mazoochi2025, Tress2024}), a high cosmic-ray ionization rate (CRIR, $\zeta \sim 10^{-15}$--$10^{-14}$~\persec{}; \citealt{Goto2013, Clark2013, Harada2015, Le2016, Oka2019}), and an intense UV radiation field ($G_0 \sim 10^{3}$--$10^{4}$; \citealt{Lis2001, Clark2013}). Collectively, these extreme conditions establish the CMZ as a natural laboratory for investigating star formation and interstellar medium processes under physical regimes unparalleled in the contemporary Galactic disk \citep{Mills2018, Henshaw2022}. 

Temperature is one of the most basic parameters to describe the physical state of gas. Under the extreme conditions of the CMZ, analyzing the temperature distribution and heating sources is essential for understanding the overall physical properties and the mechanisms governing star formation. Formaldehyde (H$_2$CO), a slightly asymmetric rotor molecule ubiquitous in the interstellar medium \citep[e.g.,][]{Downes1980, Mangum1993, Ao2013, Ginsburg2016, Yan2019, Gong2023}, is an excellent thermometer for dense molecular gas. Its fractional abundance remains roughly constant across diverse environments, including hot cores such as Orion, cloud ridges, and dark clouds \citep[e.g.,][]{Mangum1993, vanDishoeck1995, Gerin2024}. Because the relative populations of its $K$-ladders are primarily governed by collisions, ratios of emission lines from different $K$-ladders are sensitive probes of kinetic temperature ($T_{\mathrm{kin}}$) \citep{Mangum1993}.

Pioneering work by \citet{Ao2013}, using the para-H$_2$CO (p-H$_2$CO) $J=3$--$2$ transitions observed with the APEX telescope, revealed a typical gas temperature of $\sim 65$~\K{} across a $40^{\prime} \times 8^{\prime}$ region of the CMZ. This study was expanded by \citet{Ginsburg2016}, who mapped a $2^{\circ} \times 0.3^{\circ}$ area and found the warm gas to be widely distributed with significant fluctuations, reporting a widespread $T_{\mathrm{kin}}\sim 60$~\K{} component but also temperatures exceeding 100~\K{} in prominent clouds like Sgr~B2, Sgr~A, and G0.253+0.016 (The Brick). They concluded that turbulent dissipation is likely the dominant heating mechanism, with cosmic-ray heating as a possible additional contributor. \citet{Immer2016} mapped p-H$_2$CO $J=3$--$2$ and $J=4$--$3$ transitions toward seven CMZ clouds using APEX, finding that the gas ($>40$~\K{}) is significantly warmer than the dust ($\sim 25$~\K{}), and reported a correlation between line width and temperature consistent with turbulent heating. Similarly, \citet{Lu2017} conducted a high-resolution multi-line study of the 20~\kms{} cloud in the Sgr~A complex, combining SMA and APEX data, deriving gas temperatures $\gtrsim100$~\K{} and concluding that both star formation and strong turbulence likely govern the cloud's chemistry and heating. 

However, since the upper levels of the p-H$_2$CO $J=3$--$2$ transitions used as temperature diagnostics (e.g., $3_{21}$--$2_{20}$ and $3_{22}$--$2_{21}$ with $E_u/k \sim 68$~\K{}; see Table~\ref{table:transitions}) are not sufficiently high above the ground state, they cannot accurately determine kinetic temperatures significantly higher than $\sim 100$~\K{}. While the $J=4$--$3$ transitions used by \citet{Immer2016} include lines with higher upper levels ($4_{22}$--$3_{21}$ with $E_u/k \sim 82$~\K{}; Table~\ref{table:transitions}), reliable temperature measurement in the regime of reported extreme temperatures ($>100$~\K{}) requires a thermometer with even higher excitation. The ortho-H$_2$CO (o-H$_2$CO) $J=5$--$4$ transitions, particularly $5_{33}$--$4_{32}$ and $5_{32}$--$4_{31}$ with $E_u/k \sim 158$~\K{} (Table~\ref{table:transitions}), are ideally suited for this purpose. The viability of this approach was demonstrated by \citet{Mangum2019}, who successfully used the o-H$_2$CO $J=5$--$4$ transitions to constrain $T_{\mathrm{kin}}$ in the starburst galaxy NGC~253.

\begin{deluxetable}{lccccl}
\tablewidth{0pt}
\tablecaption{H$_2$CO Transitions Used in This Study\label{table:transitions}}
\tablehead{
    \colhead{Species} & \colhead{Transition} & \colhead{Frequency} & \colhead{$E_u/k$} & \colhead{Telescope} & \colhead{Notes} \\
    \colhead{} & \colhead{$J_{K_a K_c}$} & \colhead{(GHz)} & \colhead{(K)} & \colhead{} & \colhead{}
}
\startdata
p-H$_2$CO & $3_{03}$--$2_{02}$ & 218.222 & 21.0 & APEX & Main line \\
p-H$_2$CO & $3_{22}$--$2_{21}$ & 218.476 & 68.1 & APEX & Blended$^a$ \\
p-H$_2$CO & $3_{21}$--$2_{20}$ & 218.760 & 68.1 & APEX & \\
\hline
o-H$_2$CO & $5_{15}$--$4_{14}$ & 351.769 & 62.5 & JCMT & Main line \\
o-H$_2$CO & $5_{33}$--$4_{32}$ & 364.275 & 158.4 & JCMT & Blended$^b$ \\
o-H$_2$CO & $5_{32}$--$4_{31}$ & 364.289 & 158.4 & JCMT & Blended$^b$ \\
\enddata
\tablenotetext{a}{Blended with CH$_3$OH $4_{22}$--$3_{12}$; not used in this study.}
\tablenotetext{b}{These two transitions are blended into a single spectral feature and analyzed together.}
\end{deluxetable}

In this study, we present new observations of the o-H$_2$CO $J=5$--$4$ transitions toward three molecular clouds with high temperatures in the innermost region of the CMZ: The Brick (G0.253+0.016), Sgr~A1, and Sgr~A2. The Brick is one of the most massive ($>10^5$\,M$_\odot$) and dense ($>10^4$~\cmc{}) molecular clouds in the Galaxy, yet it lacks signatures of widespread star formation \citep{Longmore2012, Kauffmann2013}. Only localized star formation activity has been observed, including a small protocluster of low- to intermediate-mass protostars, with no evidence for massive protostars ($>8$\,M$_\odot$) \citep{Lis1994, Lu2021}. The cloud harbors strong solenoidal turbulence \citep{Federrath2016}, likely driven by shear in the deep gravitational potential of the CMZ \citep{Kruijssen2019, Petkova2023}. Using SMA observations of H$_2$CO $J=3$--$2$ lines, \citet{Johnston2014} identified localized high-temperature gas ($>320$~\K{}) and shock-dominated dynamics in the southern part of the cloud, suggestive of a cloud-cloud collision event. The Sgr~A1 region mainly consists of the Sticks and Stone clouds \citep{Battersby2020}, which share similar global properties but exhibit different levels of substructure complexity \citep{Battersby2020, Hatchfield2020}: the Stone cloud may be in an early stage of star formation \citep{Alboslani2025}, while the Sticks cloud remains largely quiescent. The Sgr~A2 region encompasses the 20 and 50~\kms{} clouds in the immediate vicinity of Sgr~A*. The 20~\kms{} cloud hosts active early-stage star formation with deeply embedded protostars \citep{Lu2015, Lu2017, Lu2019}, while the 50~\kms{} cloud is associated with a cluster of compact H\,{\sc ii} regions potentially triggered by the adjacent Sgr~A East supernova remnant \citep{Ho1985}. Recent ALMA observations have further revealed parsec-scale shock-driven filamentary structures at the edges of both clouds \citep{Yang2025}. The diverse physical processes and rich observational heritage of these clouds make them ideal targets for investigating the heating mechanisms of dense gas in the CMZ.

We combine the o-H$_2$CO $J=5$--$4$ data with the existing p-H$_2$CO $J=3$--$2$ data from \citet{Ginsburg2016} and analyze them with the non-local thermodynamic equilibrium (non-LTE) radiative transfer code RADEX \citep{Van2007}. This approach allows us to derive more reliable and precise measurements of the physical conditions for these clouds, and thereby clarify the roles of cosmic-ray and turbulent heating.

The structure of this paper is as follows: Section~\ref{sec:observations} introduces the JCMT observations and data reduction process. Section~\ref{sec:results} presents the temperature distribution results for the three target regions. Section~\ref{sec:discussion} discusses the origin of the high temperatures and their implications for the energy balance in the CMZ. Section~\ref{sec:conclusions} summarizes the main conclusions.

\section{Observations and Data Reduction}
\label{sec:observations}
\subsection{JCMT Observations and Data Processing}
\subsubsection{Observations}

This study observed three targets in the CMZ, ``The Brick'' (G0.253+0.016), the Sgr~A1 cloud (M0.11$-$0.08 and M0.07$-$0.08) and the Sgr~A2 cloud (20 and 50 \kms{} GMC), using the HARP receiver \citep{buckle2009} and ACSIS spectrometer on the James Clerk Maxwell Telescope (JCMT). The observations were conducted in two separate campaigns under project IDs M15AI058 (2015) and M17AP079 (2017).

We focused on observing the o-H$_2$CO $J=5$--$4$ transitions. Due to the large frequency separation, the observations were carried out in two distinct setups:

\begin{itemize}
    \item 351~\GHz{} setup: Covering the frequency range \SIrange{351.310}{352.211}{\GHz} (M15AI058) and \SIrange{350.871}{352.601}{\GHz} (M17AP079) in the signal sideband, targeting the $J=5_{15}$--$4_{14}$ transition at \SI{351.768}{\GHz}.
    \item 364~\GHz{} setup: Covering \SIrange{363.352}{365.085}{\GHz} in the signal sideband, targeting the $J=5_{33}$--$4_{32}$ and $J=5_{32}$--$4_{31}$ transitions at \SI{364.275}{\GHz} and \SI{364.289}{\GHz}, respectively.
\end{itemize}

The observational coverage was as follows:
\begin{itemize}
    \item The Brick: $J=5_{15}$--$4_{14}$ observed in 2015; $J=5_{33}$--$4_{32}$ and $J=5_{32}$--$4_{31}$ observed in 2017.
    \item Sgr~A1 and Sgr~A2: Both frequency setups observed in 2017.
\end{itemize}

Mapping observations were carried out in scan (raster) mode. The data cubes have spatial pixel scales of $\approx$7.28\arcsec{} and velocity channel widths of 0.42~\kms{}  (M15AI058) and 0.80~\kms{} (M17AP079) along the spectral axis. At the targeted frequencies, the diffraction-limited angular resolution (FWHM) of the JCMT telescope is $\approx$13\arcsec{}. The main beam efficiency ($\eta_{\mathrm{mb}}$) was taken as 0.64.

\subsubsection{Data Processing}

The reduction of the raw data was completed using the \texttt{ORAC-DR} pipeline \citep{Jenness2015}\footnote{\url{https://www.eao.hawaii.edu/oracdr/}} within the \texttt{Starlink} software package \citep{Currie2014}\footnote{\url{https://starlink.eao.hawaii.edu/starlink}}. This pipeline automatically performed standard steps, including bad channel and baseline removal, intensity calibration (converting system temperature to corrected antenna temperature $T_{\mathrm{A}}^{*}$), point source sensitivity calibration, and combination of individual scans into a final data cube in Galactic coordinates. 

The spectra observed in the 364~\GHz{} band exhibit a baseline with a characteristic fixed waveform. This feature may arise from residual atmospheric emission that is not fully removed by the standard calibration pipeline, as the Voigt-like profile is typical of atmospheric lines affected by pressure broadening. Alternatively, instrumental effects could contribute to such baseline structures. To fit and subtract this baseline, we employed a composite function model that combines two pseudo-Voigt functions (linear combinations of Gaussian and Lorentzian functions) with a quadratic polynomial background. A single pseudo-Voigt component was insufficient to reproduce the observed baseline shape, necessitating a two-component model. The specific baseline fitting process and function selection are detailed in Appendix~\ref{app:baseline}.

 After successful baseline subtraction, the achieved noise levels were $130$--$180$~\mK{} at $351.768$~\GHz{} and $50$--$80$~\mK{} at $364.103$~\GHz{} on a $T_{\mathrm{A}}^{*}$ scale with a velocity resolution of $0.8$~\kms{}.

We created masks in the baseline-subtracted data cubes at locations with significant signal, following the procedure outlined below:
\begin{itemize}
    \item To estimate the local noise level ($\mathrm{RMS}_{\mathrm{noise}}$), we computed the sample standard deviation over a 200~\kms{} velocity range devoid of detectable signal.
    \item We then smoothed the raw spectra along the velocity axis using a one-dimensional Gaussian kernel with a standard deviation of 2 channels ($\sim 1.6$~\kms{}), yielding smoothed spectra for signal identification.
    \item Given that astrophysical emission is expected to exhibit spatial and spectral continuity, we applied the following criteria to identify robust detections: (1) The intensity in the smoothed spectrum must exceed a threshold of $3 \times \mathrm{RMS}_{\mathrm{noise}}$; (2) The channels exceeding this threshold must form a contiguous segment; and (3) This segment must span at least 2 consecutive velocity channels. This continuity condition effectively suppresses spurious detections from isolated noise spikes while preserving real, spectrally coherent emission features.
\end{itemize}

\begin{figure*}[ht!]
    \centering
    \includegraphics[width=\textwidth]{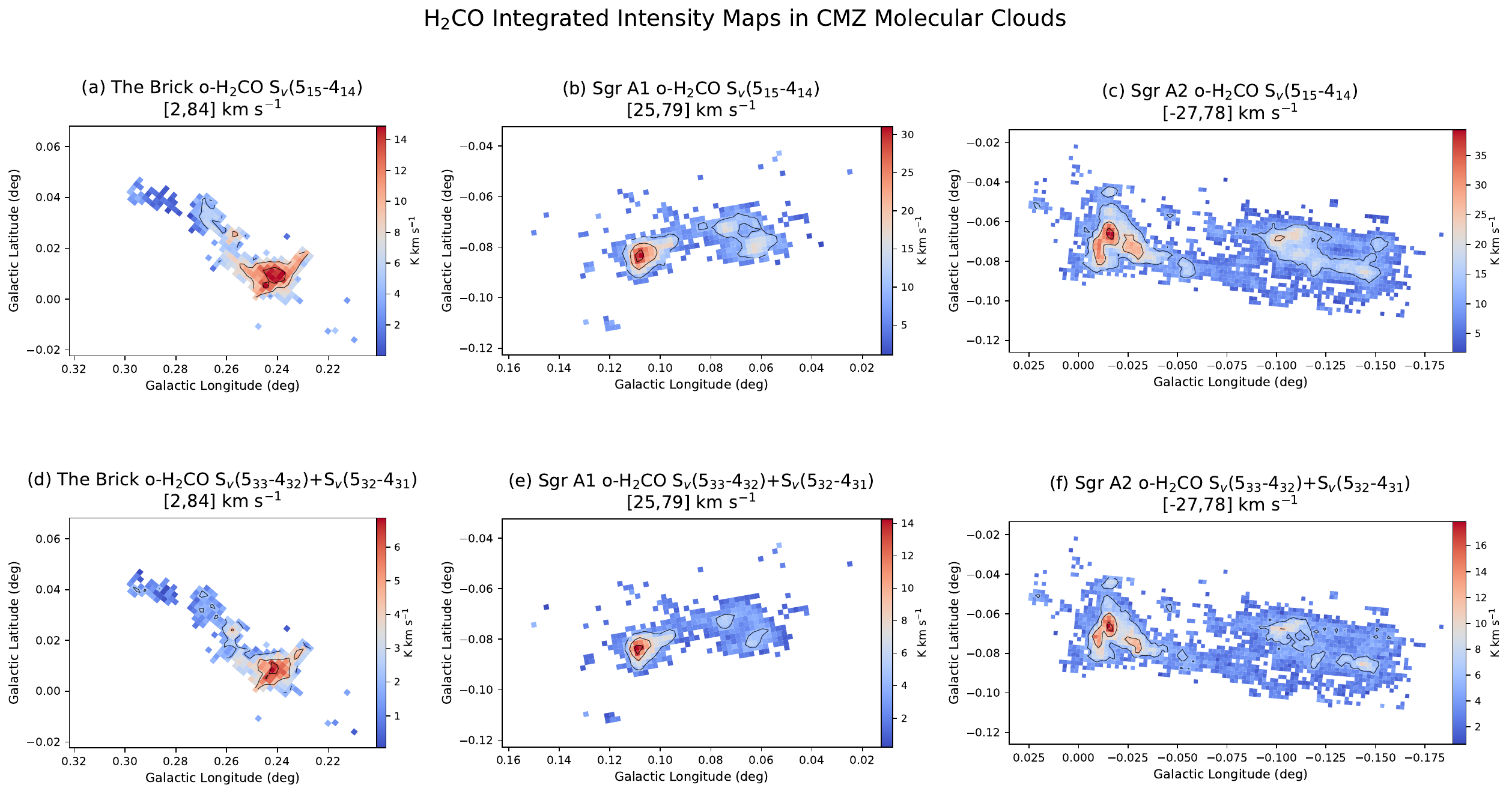}
    \caption{Integrated intensity maps of the H$_2$CO $J=5-4$ transitions for the three targeted molecular clouds after baseline subtraction and masking.
    Panels (a--c): intensity maps of the o-H$_2$CO $J=5_{15}$--$4_{14}$ transition.
    Panels (d--f): intensity maps of the combined o-H$_2$CO $J=5_{33}$--$4_{32}$ and $J=5_{32}$--$4_{31}$ transitions.
    For each subpanel, the velocity integration range is indicated above the map.
    The contours in each panel correspond to 30\%, 60\%, and 90\% of the peak integrated intensity.}
    \label{fig:integrated_intensity}
\end{figure*}

Figure~\ref{fig:integrated_intensity} shows the integrated intensity maps of the three clouds after baseline subtraction and masking.

\subsection{APEX Data}
The p-H$_2$CO transition data used in this study were taken from published APEX observations by \citet{Ginsburg2016}. This dataset covers a large area of the CMZ ($-0.4^\circ < l < 1.6^\circ$) and includes observations of p-H$_2$CO $3_{03}$--$2_{02}$, $3_{22}$--$2_{21}$, and $3_{21}$--$2_{20}$ transitions.

According to \citet{Ginsburg2016}, the telescope main beam efficiency for these data is $\eta_{\mathrm{mb}} = 0.75$, and the sensitivities of the spectra reach $\sigma = 50$--$80$~\mK{} per 1~\kms{} velocity channel and 30\arcsec{} beam on a $T_{\mathrm{A}}^{*}$ scale. For joint analysis, this study extracted the portions corresponding to the JCMT-observed regions from the aforementioned published data.

Given the difference in angular resolution between the JCMT and APEX telescopes, when comparing and jointly analyzing observational data from both telescopes, we smoothed the JCMT o-H$_2$CO $J=5$--$4$ data using a Gaussian kernel ($\mathrm{FWHM} = 26.53$\arcsec{}, approximately 3.65 pixels) to match the effective angular resolution of the APEX data ($\mathrm{FWHM} = 30$\arcsec{}).

\section{Results}
\label{sec:results}

\subsection{Ratio maps}
\label{ssec:ratio_maps}

We performed Gaussian fitting separately on the p-H$_2$CO $J=3$--$2$ lines from the APEX telescope and the o-H$_2$CO $J=5$--$4$ lines from the JCMT telescope. The fitting was conducted in two independent groups: one for the APEX data (two $J=3$--$2$ lines) and another for the JCMT data (three $J=5$--$4$ lines). We did not use the $3_{22}$--$2_{21}$ transition, which also lies near $\sim$218~\GHz{}, because it is closely spaced in frequency to the CH$_3$OH $4_{22}$--$3_{12}$ line and is likely severely blended (\citealt{Ginsburg2016}). Within each group, we assumed that all lines share the same line-of-sight velocity ($v_{\mathrm{lsr}}$) and velocity dispersion ($\sigma_v$), given their similar excitation conditions. This approach is justified by the results: the velocity fields derived from the two groups are largely consistent.

We defined two key line intensity ratios as our temperature diagnostics. For the $J=3$--$2$ data, we used the ratio:
\begin{equation}
R_{3-2} = \frac{S_v(3_{21}\text{-}2_{20})}{S_v(3_{03}\text{-}2_{02})}
\end{equation}
 For the $J=5$--$4$ data, the $5_{33}$--$4_{32}$ and $5_{32}$--$4_{31}$ transitions are separated by only $\sim$14~\MHz{} ($\sim$11.5~\kms{}) and are blended into a single spectral feature. Since both lines are optically thin ($\tau \ll 1$) and their intensities add linearly, we treat them as a combined feature and define the ratio as:
\begin{equation}
R_{5-4} = \frac{S_v(5_{33}\text{-}4_{32}) + S_v(5_{32}\text{-}4_{31})}{S_v(5_{15}\text{-}4_{14})}
\end{equation}

For most spectra exhibiting a single-peaked profile, we employed a single-component Gaussian model for fitting. For spectra showing a double-peaked structure, we used a two-component Gaussian model for fitting. The $J=3$--$2$ and $J=5$--$4$ data exhibit consistent single- or double-peaked morphology at the vast majority of spatial pixels. This consistency indicates that the observed multi-peaked structures primarily originate from multiple velocity components in the gas itself, rather than self-absorption due to high optical depth. Since the $3_{03}$--$2_{02}$ line has the highest intensity, we used this line to decide whether to apply two-component Gaussian fitting. We first performed single-component Gaussian fits to all lines and calculated the residuals. If significant signals exceeding $3\sigma$ were present in the smoothed residuals, the spectrum was judged to have multiple velocity components, and subsequently, a two-component Gaussian model was used to refit all lines at that position.

The parameterization of the single-Gaussian model is as follows:
\begin{itemize}
 	\item For the $J=3$--$2$ line group, the model contains 4 free parameters: the line centroid velocity $v_{\mathrm{lsr}}$, the velocity dispersion $\sigma_v$, the amplitude $A$ of the $3_{03}$--$2_{02}$ main line, and the intensity ratio $R_{3-2}$.
	\item For the $J=5$--$4$ line group, the model contains 5 free parameters: $v_{\mathrm{lsr}}$, $\sigma_v$, the amplitude $A$ of the main line $5_{15}$--$4_{14}$, and the peak intensity ratios $r_1$ and $r_2$ of the two high-$K$ lines relative to the main line. Due to line blending, $r_1$ and $r_2$ cannot be independently assigned to specific transitions, but their sum $r_1 + r_2$ is physically equivalent to the required integrated intensity ratio $R_{5-4}$, which was used in the subsequent analysis.
\end{itemize}
The corresponding double-Gaussian model employs two independent sets of these parameters. This means each velocity component ($i=1,\,2$) is fitted independently using the same number of parameters as defined for the single-Gaussian model. For spatial pixels fitted with the double-Gaussian model, we took the average of the corresponding parameters from each component as the final representative value for that pixel.

\begin{figure*}[ht!]
    \centering
    \includegraphics[width=\textwidth]{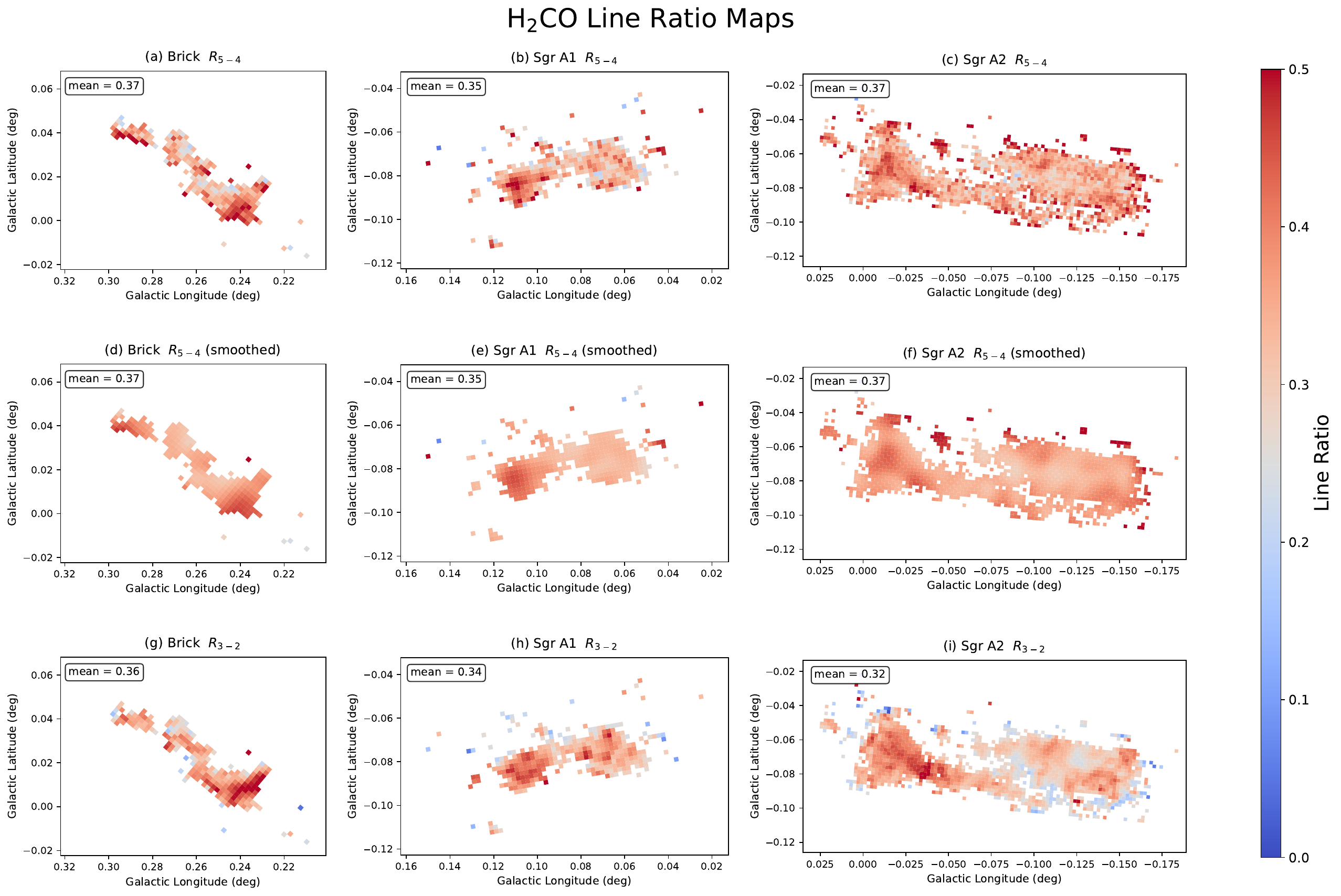}
    \caption{
        Spatial distribution of the o-H$_2$CO line intensity ratios, $R$, for the three target molecular clouds.
        Top row (a--c): Ratio $R_{5-4}$ derived from the original JCMT data for The Brick, Sgr~A1, and Sgr~A2, respectively.
        Middle row (d--f): Ratio $R_{5-4}$ for the same clouds after smoothing the JCMT data to the APEX resolution.
        Bottom row (g--i): Ratio $R_{3-2}$ from the APEX data \citep{Ginsburg2016}.
        The color bar indicates the value of the respective ratio, with the scale restricted to the range 0--0.5 for uniform comparison across all panels.}
    \label{fig:ratio_distribution}
\end{figure*}

Using the \texttt{pyspeckit} \citep{Ginsburg2011, Ginsburg2022}\footnote{\url{https://pyspeckit.readthedocs.io/en/latest/}} package, we performed the above fitting on the original JCMT $J=5$--$4$ data, the JCMT $J=5$--$4$ data smoothed to the APEX resolution (30\arcsec{}), and the APEX $J=3$--$2$ data, extracting the $R_{3-2}$ and $R_{5-4}$ values for each pixel. Figure~\ref{fig:ratio_distribution} presents the spatial distribution of these ratios across the three target molecular clouds.

\subsection{Non-LTE Modeling and Temperature Constraints based on RADEX}
To reliably derive the gas kinetic temperature ($T_{\mathrm{kin}}$) from the observed H$_2$CO spectral line intensities, we employed the non-LTE radiative transfer program \texttt{RADEX} \citep{Van2007}. \texttt{RADEX} solves the statistical equilibrium and radiative transfer equations using an escape probability formalism with various geometrical assumptions. We adopted the Large Velocity Gradient (LVG) approximation \citep{Sobolev1960}, which is well-suited to turbulent molecular clouds with significant internal velocity fields, such as those in the CMZ. The program numerically predicts molecular line intensities and brightness temperatures for given physical parameters such as H$_2$ number density ($n_{\mathrm{H_2}}$), kinetic temperature ($T_{\mathrm{kin}}$), molecular column density ($N_{\mathrm{molecule}}$), and velocity gradient ($dv/dr$).

\subsubsection{Model Grid Construction}
We utilized the \texttt{myRadex} optimization software \citep{myRadex}\footnote{\url{https://github.com/fjdu/myRadex}} to construct a theoretical spectral line grid covering a wide parameter space. Collisional excitation rate coefficients for H$_2$CO were taken from the LAMDA database \citep{van2020}, specifically adopting the rate coefficients from \citet{Wiesenfeld2013}. We used p-H$_2$ as the collisional partner. The model grid covered the following parameter ranges:

1. H$_2$ number density ($n_{\mathrm{H_2}}$): 20 points, logarithmically sampled, ranging from $10^{2.5}$ to $10^{7}$~\cmc.

2. Para-H$_2$CO column density ($N$(p-H$_2$CO)): 30 points, logarithmically sampled, ranging from $10^{11}$ to $10^{15}$~\cms. For the o-H$_2$CO/p-H$_2$CO ratio, see Section~\ref{ssec:density_opr}.

3. Kinetic temperature ($T_{\mathrm{kin}}$): 50 points, linearly sampled, ranging from 10 to 350~\K{}.

4. Velocity gradient ($dv/dr$): fixed at 5~\kmspc{}, a typical value for Galactic center molecular clouds based on estimates ranging from 3 to 6~\kmspc{} by \citet{Dahmen1998} and adopted in \citet{Ao2013}.

To improve the precision of parameter estimation, we performed three-dimensional (3D) spline interpolation on the initial grid, increasing the grid to a total of $250 \times 100 \times 150$ (density $\times$ column density $\times$ temperature) points.

\subsubsection{Kinetic Temperature Maps}
We compared model predictions with the following observational results to constrain the temperature:

1. Line Intensity Ratio Constraints: We used three methods to constrain the kinetic temperature of the molecular clouds. These included (a) using only the integrated intensity ratio $R_{3-2}$ from the APEX p-H$_2$CO $J=3$--$2$ data; (b) using only the integrated intensity ratio $R_{5-4}$ from the JCMT o-H$_2$CO $J=5$--$4$ data; and (c) using both $R_{3-2}$ and $R_{5-4}$ as joint constraints. For the joint analysis, the JCMT data were smoothed to match the APEX resolution.

2. Brightness Temperature Lower Limit Constraints: To ensure model rationality, we required that the predicted spectral line brightness temperature ($T_B$) from the model be no lower than the observed value.

3. Column Density and Abundance Constraints: For each pixel in our data, we assigned the H$_2$ column density from the map of \citet{Battersby2025a} using the nearest-neighbor value at the corresponding coordinates, with a characteristic uncertainty of a factor of 10 ($\pm$1 in log space). The resulting mean H$_2$ column densities are $1.61 \times 10^{23}$ \cms{} for the Brick, $9.78 \times 10^{22}$ cm$^{-2}$ for Sgr~A1, and $1.46 \times 10^{23}$ \cms{} for Sgr~A2. For the H$_2$CO abundance, we adopted $X(\mathrm{p-H}_2\mathrm{CO}) = N(\mathrm{p-H}_2\mathrm{CO})/N(\mathrm{H}_2) = 10^{-9.08 \pm 1}$, allowing for more than two orders of magnitude variation in the H$_2$CO abundance, consistent with previous studies \citep[e.g.,][]{Ginsburg2016, Carey1998, Wootten1978, Mundy1987}. For the ortho-to-para ratio (OPR), defined as $N(\mathrm{o-H}_2\mathrm{CO}) / N(\mathrm{p-H}_2\mathrm{CO})$, we first systematically tested fixed values from 1 to 3. These tests revealed that the derived kinetic temperature is largely insensitive to the specific OPR choice. To further validate this finding, we subsequently treated the OPR as a free parameter in our radiative transfer modeling. The resulting OPR values, derived from the joint analysis of both transitions, converge to a mean of approximately 3 across the target molecular clouds (see Section~\ref{ssec:density_opr}). This empirical confirmation supports our adoption of OPR $= 3$ (the statistical equilibrium value under thermal conditions; \citealt{Kahane1984, Troscompt2009}) for the subsequent analysis.

For each point in the parameter space, we calculated its $\chi^2$ statistic against all observational constraints. The total $\chi^2$ was obtained by summing the $\chi^2$ values from individual constraints. We converted this into a joint likelihood function: $\mathcal{L}(n_{\mathrm{H_2}}, T_{\mathrm{kin}}, N_{\mathrm{p-H}_2\mathrm{CO}}) \propto \exp(-\chi_{\mathrm{total}}^2 / 2)$. Finally, the kinetic temperature for each pixel was determined by performing a likelihood-weighted average over the entire three-dimensional parameter space. To assess the quality of these estimates, we also computed the uncertainty for each pixel, which represents the measurement precision at a single spatial position. For each pixel, after normalizing the likelihood to unit total probability, we identified the $1\sigma$ (68.3\% confidence) highest-density credible region by finding a likelihood threshold such that the cumulative probability of all grid points exceeding that threshold equals 0.683. The kinetic temperature uncertainty is then defined as half the full range of temperatures spanned by this region, i.e., $(T_{\mathrm{max}} - T_{\mathrm{min}})/2$.

\subsubsection{Results of Kinetic Temperature Constraints}
\label{ssec:temp_results}

\begin{figure*}[h!]
    \centering
\includegraphics[width=0.95\textwidth]{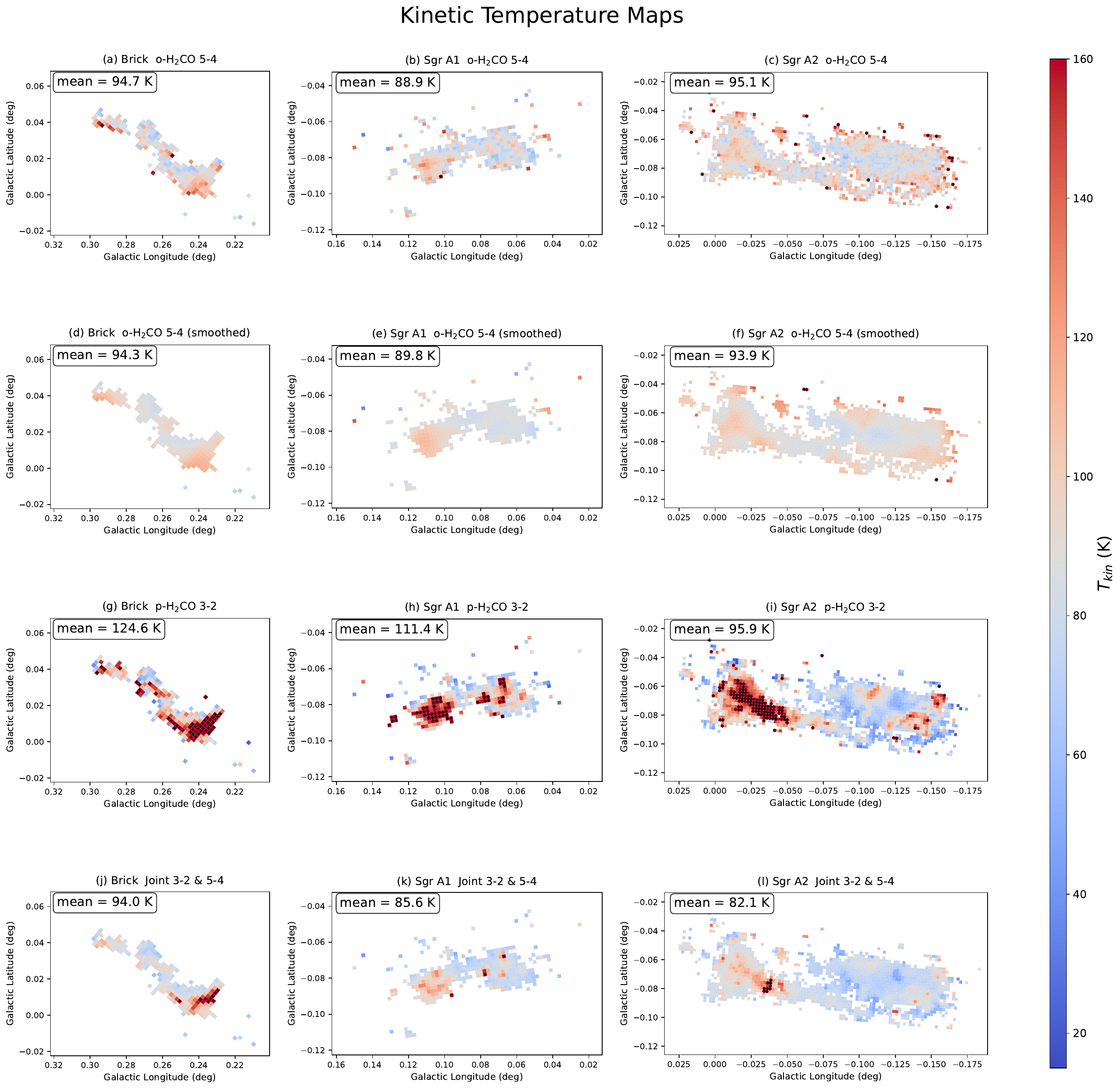}
    \caption{
        Distribution of the kinetic temperature ($T_{\mathrm{kin}}$) across the three target molecular clouds under different observational constraints.
        The color of each pixel corresponds to the kinetic temperature, as indicated by the shared color bar (right), with values restricted to an upper limit of 160~\K{} where the majority of the data lie.
        Pixels with temperatures exceeding 160~\K{} are marked with black plus signs (+).
        Columns, from left to right, represent: The Brick (G0.253+0.016), Sgr~A1, and Sgr~A2.
        Rows, from top to bottom, show temperatures derived from: the original JCMT $J=5$--$4$ data alone; the smoothed JCMT $J=5$--$4$ data alone; the APEX $J=3$--$2$ data alone \citep{Ginsburg2016}; and the joint constraints using both smoothed JCMT $J=5$--$4$ and APEX $J=3$--$2$ data.
    }
    \label{fig:temperature_distribution}
\end{figure*}

Based on observations of the H$_2$CO $J=3$--$2$ and $J=5$--$4$ lines and non-LTE radiative transfer modeling, we have derived the kinetic temperature ($T_{\mathrm{kin}}$) distribution for the three target molecular clouds (The Brick, Sgr~A1, and Sgr~A2) in the CMZ. Figure~\ref{fig:temperature_distribution} presents the resulting temperature maps derived using different constraints: JCMT $J=5$--$4$ data alone, smoothed JCMT $J=5$--$4$ data alone, APEX $J=3$--$2$ data alone, and the joint constraint from both datasets. The results shown in this figure correspond to an OPR value of 3. 

Our analysis shows that relying on a single group of transitions for temperature measurement introduces systematic bias, but its manifestation and physical cause differ. When using only the H$_2$CO $J=3$--$2$ lines, significant temperature overestimation occurs in high kinetic temperature regions ($T_{\mathrm{kin}} > 100$~\K{}). This overestimation is most severe in the hottest areas, and for gas at the same true temperature, the degree of overestimation increases with increasing gas column density.
\begin{figure*}[ht!]
\centering
\includegraphics[width=\textwidth]{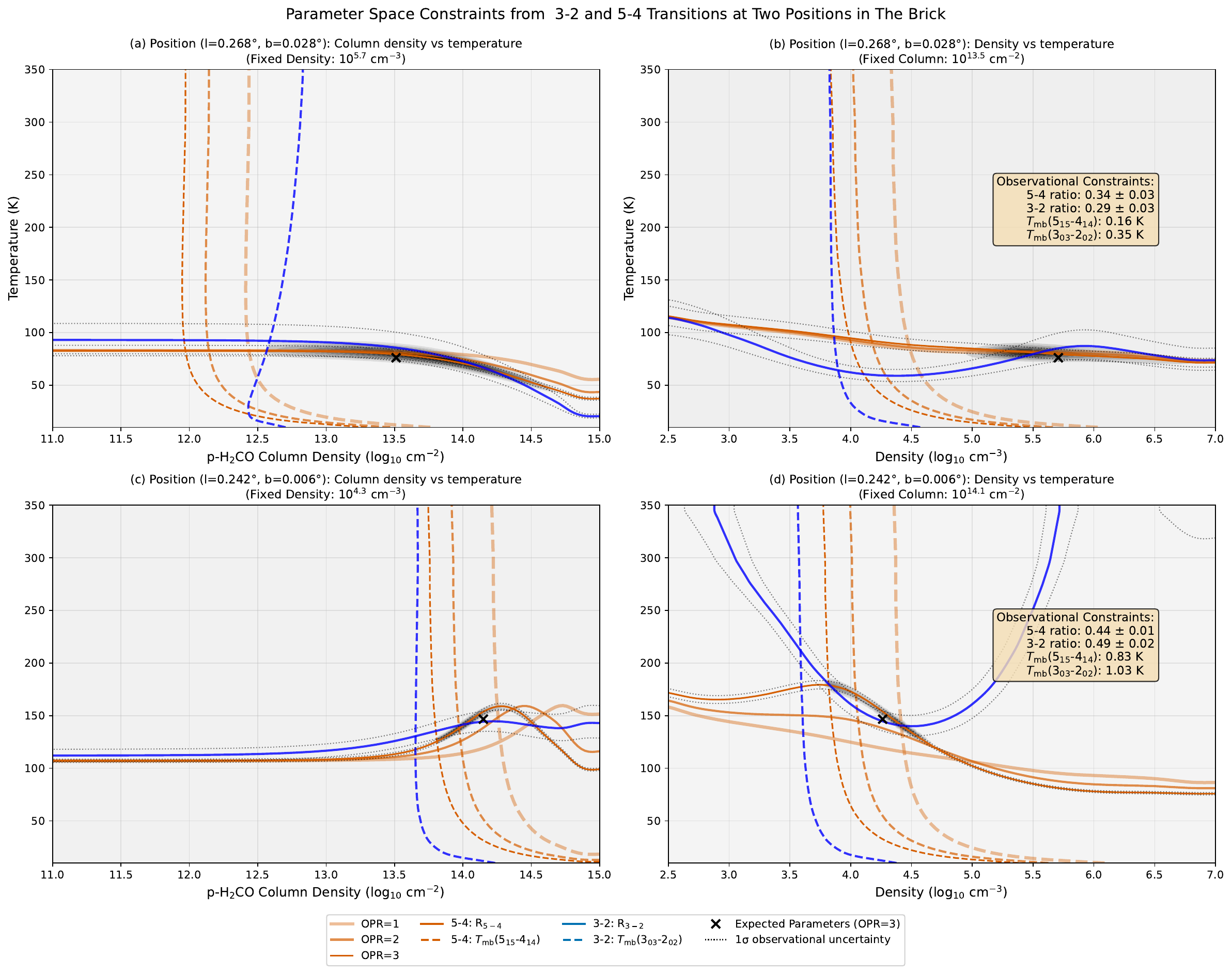}
\caption{
Parameter space analysis for two selected positions in The Brick (G0.253+0.016) at ($l = 0.268^\circ$, $b = 0.028^\circ$) (top row) and ($l = 0.242^\circ$, $b = 0.006^\circ$) (bottom row).
Panels (a) and (c) show column density--temperature slices at fixed H$_2$ densities (indicated above each panel), while panels (b) and (d) show density--temperature slices at fixed p-H$_2$CO column densities (indicated above each panel).
The grayscale background represents the likelihood distribution from the joint constraints of the $J=3$--$2$ and $J=5$--$4$ transitions at $\mathrm{OPR} = 3$, with darker shades indicating higher likelihood.
Colored contours show model predictions for different OPR values (blue: $J=3$--$2$ constraints; red: $J=5$--$4$ constraints). Line styles distinguish the type of constraint (solid: line ratios; dashed: line intensities). For clarity, only constraints from the $3_{03}$--$2_{02}$ and $5_{15}$--$4_{14}$ transitions are visualized here; the full likelihood distribution, however, incorporates additional constraints from the $3_{21}$--$2_{20}$ and $5_{33}$--$4_{32}$ \& $5_{32}$--$4_{31}$ transitions. Contour lines for different OPR values are distinguished by color intensity, ranging from light ($\mathrm{OPR} = 1$) to bold ($\mathrm{OPR} = 3$).
Black dashed lines indicate the $1\sigma$ observational uncertainty on the line ratios, and black ``$\times$'' symbols mark the expected parameter values for OPR = 3.
Observed line ratios and main-beam temperatures for each position are provided by the panels on the right hand side.
}
\label{fig:likelihood_examples}
\end{figure*}

Figure~\ref{fig:likelihood_examples} shows two-dimensional slices through the parameter space for representative high- and low-temperature regions, illustrating the constraints in the $n_{\mathrm{H_2}}$--$T_{\mathrm{kin}}$ and $N_{\mathrm{p-H}_2\mathrm{CO}}$--$T_{\mathrm{kin}}$ planes. The figure displays model predictions for different OPR ($\mathrm{OPR} = 1,\, 2,\, 3$) as colored contours, with line styles distinguishing between line ratio constraints and primary line intensities. 

The systematic bias of temperature overestimation when using a single line ratio ($R_{3-2}$) originates from the fundamental behavior of level populations as a function of temperature. As elucidated by \citet{Mangum1993}, at low temperatures ($T_{\mathrm{kin}} \lesssim 70$~\K{}), the upper levels of the $J=3$--$2$ transitions (e.g., $3_{21}$--$2_{20}$, $E_u/k \sim 68$~\K{}) are subthermally populated. Consequently, the line ratio $R_{3-2}$ is highly sensitive to temperature and increases steeply as temperature rises. However, at sufficiently high temperatures ($T_{\mathrm{kin}} \gtrsim 100$~\K{}), these levels approach thermalization. The ratio $R_{3-2}$ then asymptotically approaches a limiting value set by statistical weights, becoming nearly independent of temperature.

For gas temperatures above 100~\K{}, this physical behavior results in a highly asymmetric likelihood function when using an observed $R_{3-2}$ value. Possibilities for temperatures below the true value are sharply excluded, as they would produce a distinctly lower line ratio. In contrast, possibilities for temperatures above the true value produce ratios similar to the true value due to thermalization, creating an extended tail in the likelihood function. Consequently, this inference systematically biases the derived mean temperature upward. In our study, the regions previously reported with extreme temperatures ($>100$~\K{}) based solely on $J=3$--$2$ data \citep[e.g.,][]{Ginsburg2016} are often characterized by higher density and column density. These conditions exacerbate the described systematic bias, causing these regions to appear as statistical outliers rather than indicating a true extreme deviation in kinetic temperature.

Conversely, when only the H$_2$CO $J=5$--$4$ lines are used, temperature overestimation occurs primarily in low-density regions. This is because $J=5$--$4$ emissions require higher excitation. At relatively low column densities, the $J=5$--$4$ lines exhibit diminished intensity and a compromised signal-to-noise ratio, leading to increased uncertainties in the measured line ratios. Moreover, in high-temperature regions, using only the $J=5$--$4$ data can lead to an underestimation of temperature. This occurs because at elevated temperatures the line ratio $R_{5-4}$ becomes sensitive to the gas density and decreases monotonically with increasing $n_{\mathrm{H_2}}$ above $\sim 10^{4}$~\cmc{}. In the absence of additional density constraints, the observed line ratio can be satisfied by a family of solutions that includes combinations of higher density and correspondingly lower temperature, leading to systematic underestimation of $T_{\mathrm{kin}}$.

\begin{deluxetable}{lcccccc}
    \tablewidth{\columnwidth}
    \tabletypesize{\scriptsize}
    \tablecaption{Kinetic temperatures and uncertainties under different constraints\label{table:temperature_results}}
    \tablehead{
        \colhead{Method} & \colhead{Brick} & \colhead{} & \colhead{Sgr~A1} & \colhead{} & \colhead{Sgr~A2} & \colhead{} \\
        \colhead{} & \colhead{Avg. (K)} & \colhead{Err. (K)} & \colhead{Avg. (K)} & \colhead{Err. (K)} & \colhead{Avg. (K)} & \colhead{Err. (K)}
    }
    \startdata
        3--2 & 124.6 & 75.9 & 111.4 & 62.8 & 95.9 & 51.6 \\
        5--4 & 94.7 & 60.1 & 88.9 & 52.0 & 95.1 & 58.7 \\
        5--4 (smoothed) & 94.3 & 55.4 & 89.8 & 48.6 & 93.9 & 54.6 \\
        joint (OPR$=1$) & 91.0 & 24.0 & 83.8 & 18.3 & 84.7 & 16.1 \\
        joint (OPR$=2$) & 93.3 & 31.6 & 85.3 & 24.5 & 83.9 & 23.7 \\
        joint (OPR$=3$) & 94.0 & 33.8 & 85.6 & 25.2 & 82.1 & 25.1 \\
    \enddata
\tablenotetext{a}{"Err" values are the mean 1$\sigma$ measurement errors per pixel.}    
\tablenotetext{b}{Joint constraints using both smoothed JCMT $J=5$--$4$ and APEX $J=3$--$2$ data.}
\end{deluxetable}

Utilizing information from both the $J=3$--$2$ and $J=5$--$4$ lines effectively overcomes the limitations of the individual datasets. For high-temperature regions in particular, the joint analysis provides good constraints on density, yielding a more reliable kinetic temperature distribution. However, when both transitions are used jointly, we must consider the influence of an uncertain OPR on the derived kinetic temperatures. We note that the temperature estimates derived from either the $J=3$--$2$ or the $J=5$--$4$ data alone are completely independent of the OPR, as each set of transitions involves only one spin isomer (para- or ortho-H$_2$CO). In contrast, when the two sets are combined, the constraints become sensitive to the OPR. We therefore tested the impact of varying the OPR between 1.0 and 3.0 on the results. Table~\ref{table:temperature_results} lists the average temperatures and their associated uncertainties for the three molecular clouds derived from the joint constraints under different OPR assumptions. The ``Err'' values listed in the table represent the mean $1\sigma$ measurement uncertainties per pixel, averaged over all valid pixels in each cloud. These values reflect the typical precision of our temperature estimates at a single spatial position, rather than the intrinsic temperature dispersion within the cloud; the calculation method is detailed in Section~\ref{ssec:temp_results}. The test shows that after smoothing the JCMT data to match the resolution of the APEX data, the resulting temperature distribution from the joint constraints is insensitive to the specific value of the OPR; the temperature differences obtained under different OPR assumptions are much smaller than the uncertainties introduced by other model parameters. This result indicates that, provided the resolution and signal-to-noise ratio are properly handled, the joint analysis method is robust against the OPR. Additionally, we attempted to constrain the OPR for the three clouds; related results will be discussed in detail in the next subsection.

The results from the joint constraints show that the average temperatures in the previously reported high-temperature ($> 100$~\K{}) regions within the three molecular clouds are all revised to below 100~\K{}. While these temperatures remain at the high end of the temperature distribution for the CMZ, they are no longer extreme outliers. It is noteworthy that the high-temperature regions identified by the joint analysis show significant spatial overlap with the high-temperature regions indicated by the respective single-constraint methods. This confirms the reality of the high-temperature structures. The joint constraints do not simply lower the temperature in all regions but provide a more precise constraint on the physical conditions, thereby systematically correcting the absolute temperature values deduced from the $J=3$--$2$ transitions alone.

\subsubsection{Constraining Gas Density and Ortho-to-Para Ratio}
\label{ssec:density_opr}
\begin{figure*}[ht!]
    \centering
    \includegraphics[width=0.95\textwidth]{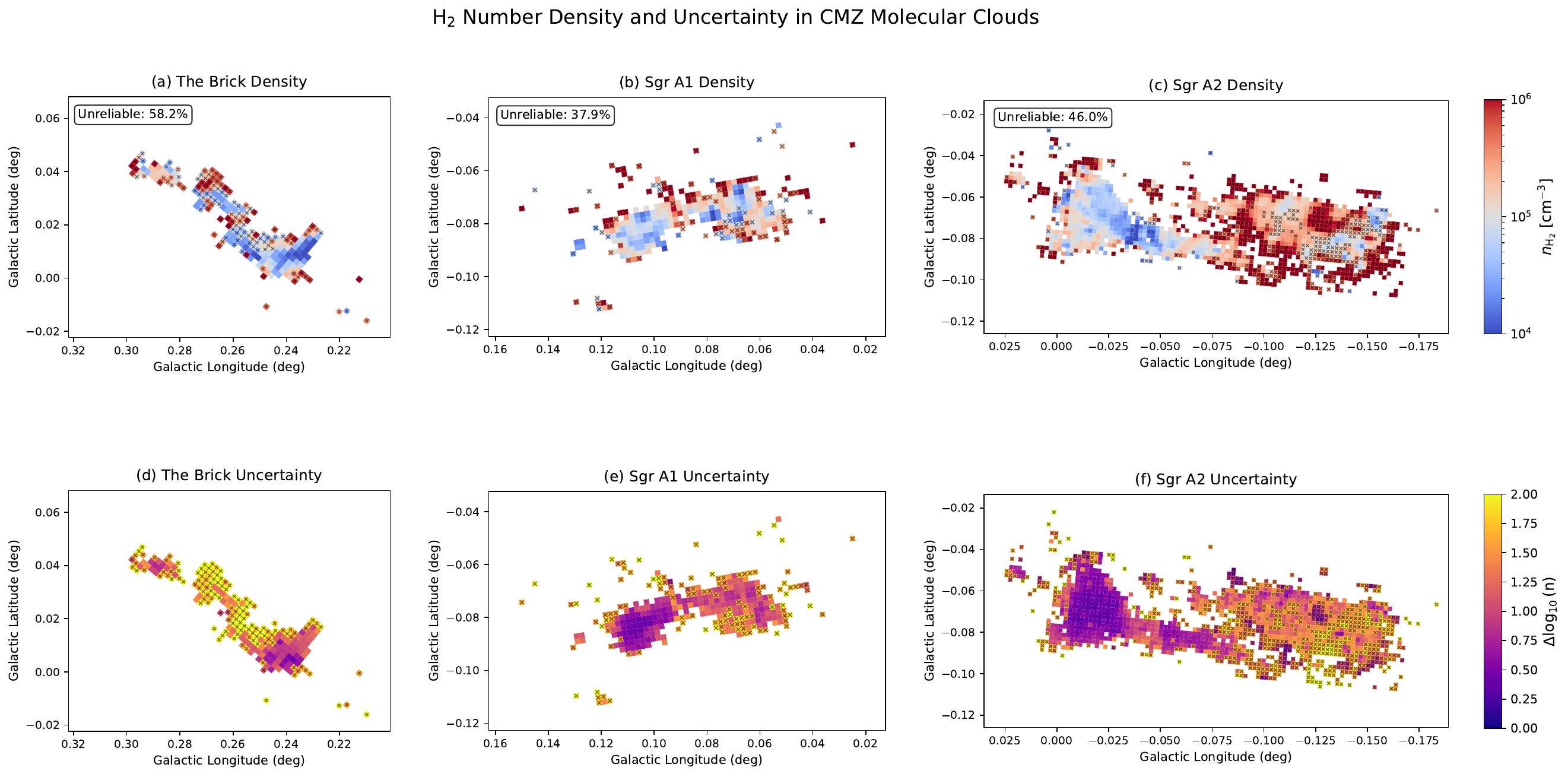}
    \caption{
        Spatial distribution of the molecular hydrogen number density ($n_{\mathrm{H_2}}$) and its uncertainty for the three target molecular clouds. Densities are derived from joint constraints of the H$_2$CO $J=3$--$2$ and $J=5$--$4$ transitions, assuming $\mathrm{OPR} = 3$.
        Upper panels (a--c): Density maps of The Brick (G0.253+0.016), Sgr~A1, and Sgr~A2 (left to right), displayed on a logarithmic scale.
        Lower panels (d--f): Corresponding uncertainty maps ($\Delta\log_{10}[n(\mathrm{H_2})]$) for each cloud.
        The color bars on the right indicate the density (upper panels) and uncertainty (lower panels) values.
        Pixels with unreliable density estimates ($\Delta\log_{10}[n(\mathrm{H_2})] > 1.5$ or the upper bound reaching the prior boundary at $\log_{10}[n(\mathrm{H_2})] = 7$) are marked with ``x''. The label in the upper-left corner of each density panel gives the percentage of pixels classified as unreliable.
       }
    \label{fig:density_map}
\end{figure*}

Through the joint analysis of both transitions, we derived constraints on the kinetic temperature and simultaneously computed the likelihood-weighted average H$_2$ number density across the parameter space grid. Figure~\ref{fig:density_map} presents the density distribution and its 1$\sigma$ uncertainty range derived from the joint constraints of the two sets of transitions under the assumption of $\mathrm{OPR} = 3$.

By examining the size of the uncertainty range, it is evident that pixels in high-temperature regions yield significantly more reliable density estimates compared to the vast majority of pixels in low-temperature regions. This behavior is expected: at lower temperatures, both the $J=3$--$2$ and $J=5$--$4$ line ratios exhibit weak sensitivity to variations in density, resulting in a broad, nearly flat likelihood distribution along the density axis (as illustrated in Fig.~\ref{fig:likelihood_examples}), which precludes effective constraints. Conversely, in high-temperature regions, the line ratios are not only sensitive to density changes but also differ in their trends, thereby jointly confining the density to a narrower range and providing significantly better constraints. We adopt two criteria for ``well-constrained'' density estimates: (1) the width of the $1\sigma$ credible interval satisfies $\Delta\log_{10}[n(\mathrm{H}_2)] < 1.5$, and (2) the upper bound of the 1$\sigma$ credible interval does not reach the prior boundary at $\log_{10}[n(\mathrm{H}_2)] = 7$, ensuring that the density is constrained from above rather than merely providing a lower limit. The fractions of pixels meeting these criteria are 41.8\% for the Brick, 62.1\% for Sgr~A1, and 54.0\% for Sgr~A2.

\begin{deluxetable}{lcccccc}
    \tablewidth{\columnwidth}
    \tabletypesize{\scriptsize}
    \tablecaption{Number densities and their uncertainties under different constraints\label{table:density_results}}
    \tablehead{
        \colhead{Method} & \multicolumn{2}{c}{Brick} & \multicolumn{2}{c}{Sgr~A1} & \multicolumn{2}{c}{Sgr~A2} \\
        \colhead{} & \colhead{Avg.} & \colhead{Err.} & \colhead{Avg.} & \colhead{Err.} & \colhead{Avg.} & \colhead{Err.}
    }
    \startdata
        3--2 & 5.54 & 1.51 & 5.60 & 1.47 & 5.58 & 1.47 \\
        5--4 & 5.22 & 1.46 & 5.26 & 1.41 & 5.25 & 1.43 \\
        5--4 (smoothed) & 5.15 & 1.49 & 5.22 & 1.42 & 5.22 & 1.44 \\
        joint (OPR$=1$) & 4.84 & 0.81 & 5.15 & 0.81 & 5.28 & 0.75 \\
        joint (OPR$=2$) & 4.68 & 0.65 & 4.96 & 0.62 & 5.11 & 0.58 \\
        joint (OPR$=3$) & 4.61 & 0.52 & 4.89 & 0.46 & 5.04 & 0.45 \\
    \enddata
\tablenotetext{a}{"Err" values are the mean 1$\sigma$ measurement errors per pixel.}
\tablenotetext{b}{Joint constraints using both smoothed JCMT $J=5$--$4$ and APEX $J=3$--$2$ data.}
\tablenotetext{c}{All density values are logarithms (base 10) of the number density in units of cm$^{-3}$, i.e., $\log_{10}[n(\mathrm{H}_2)/\mathrm{cm}^{-3}]$.}
\end{deluxetable}

Table~\ref{table:density_results} summarizes the statistical properties of the well-constrained density values. To ensure a fair comparison across different OPR assumptions, we used the same set of spatial pixels that are well-constrained under the $\mathrm{OPR} = 3$ assumption (i.e., those satisfying $\Delta\log_{10}[n(\mathrm{H}_2)] < 1.5$ and with the upper bound of the $1\sigma$ credible interval below the prior boundary at $\log_{10}[n(\mathrm{H}_2)] = 7$, as shown in Figure~\ref{fig:density_map}). For these pixels, we present the density results for OPR values ranging from 1.0 to 3.0. The assumed OPR has a more pronounced impact on the derived density than on the temperature, indicating that our density estimates are subject to considerable uncertainty. Nevertheless, we find that for all three molecular clouds and across the tested OPR range, the densities lie in the range $n_{\mathrm{H_2}} \sim 10^{4.5}$--$10^{5.5}$~\cmc{}, which is consistent with findings from other studies \citep{Immer2016}.

Building upon the density constraints, we investigated the OPR of H$_2$CO across the three target clouds. The OPR provides valuable insights into the formation history and thermal evolution of formaldehyde \citep{Kahane1984}. In warm molecular clouds like Orion, \citet{Kahane1984} found an OPR consistent with the statistical equilibrium value of 3, suggesting gas-phase formation or thermal equilibrium at temperatures above $\sim$40~\K{}. In contrast, cold dark clouds such as TMC1 and L183 exhibit significantly lower OPR values ($\sim 1$--$2$), indicative of formation on cold grain surfaces where nuclear spin states thermalize to the low dust temperature ($\sim 10$--$15$~\K{}). The energy difference between the ground states of ortho- and para-H$_2$CO ($\sim$15~\K{}) is smaller than that of H$_2$, making the OPR a sensitive probe of the thermal conditions during molecule formation.

The methodology for constraining the OPR follows an approach similar to that presented in \citet{Mazumdar2022}. For each pixel, we determined the OPR via a $\chi^2$ constraint that simultaneously fits the observed intensities of the o-H$_2$CO and p-H$_2$CO transitions. This process is performed within a two-dimensional parameter space of H$_2$ number density ($n_{\mathrm{H_2}}$) and species column density ($N$), under the fixed kinetic temperature ($T_{\mathrm{kin}}$) derived from our joint analysis.

Specifically, for the o-H$_2$CO and p-H$_2$CO transitions separately, we compute the confidence regions in the $n_{\mathrm{H_2}}$--$N_{\mathrm{H_2CO}}$ plane that are consistent with the observed peak line intensities. The OPR is then estimated as the ratio of the two independently constrained column densities at a chosen density. For pixels where the density is well-constrained (i.e., $\Delta\log_{10}[n(\mathrm{H_2})] < 1.5$ and the upper bound not reaching the prior boundary), we use the derived density value from the joint analysis. For pixels with poorly constrained density, we adopt a characteristic density of $n_{\mathrm{H_2}} = 10^5$~\cmc{}, based on the typical value from well-constrained regions.

\begin{figure}[ht!]
    \centering
    \includegraphics[width=\columnwidth]{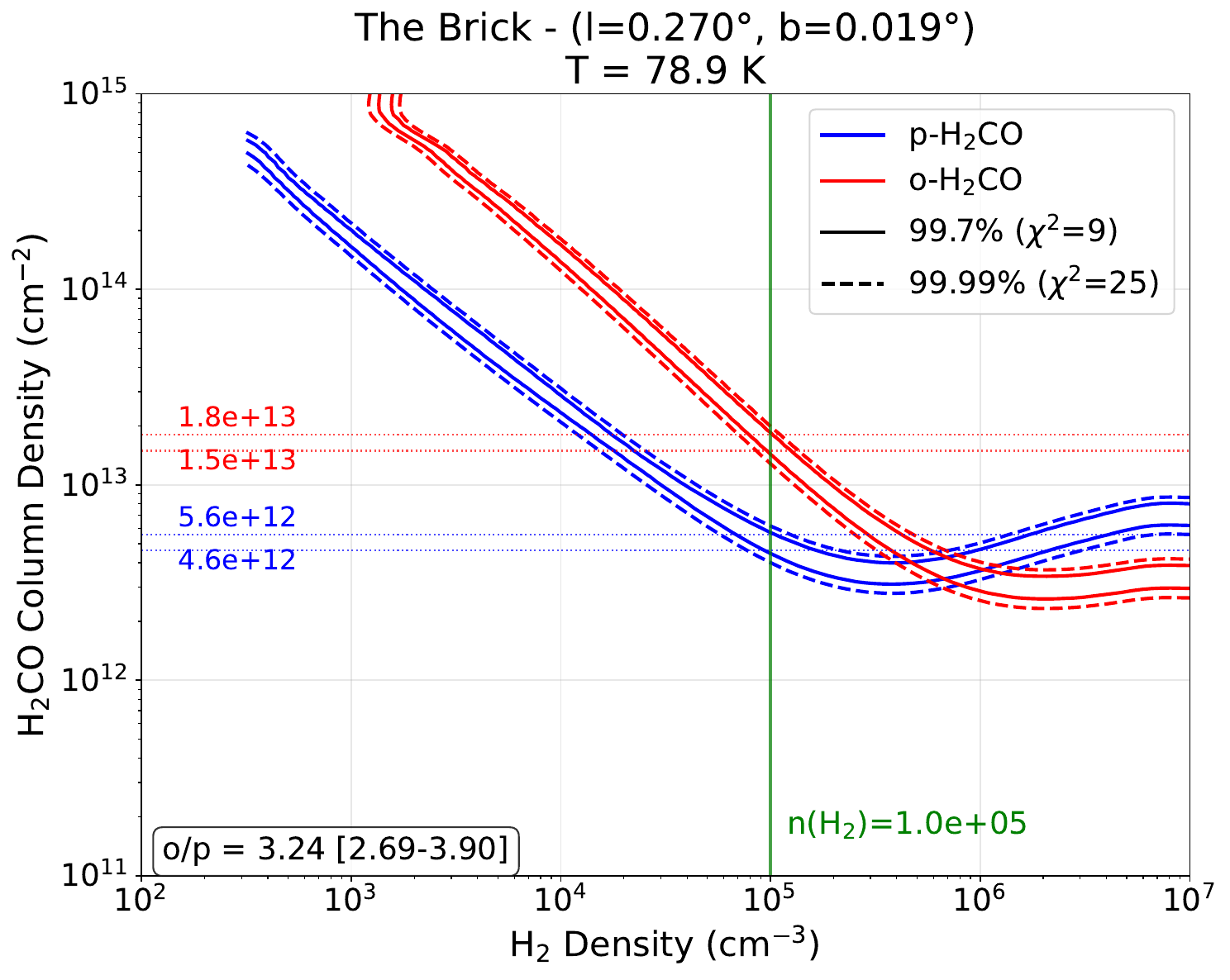}
    \caption{
    Constraints on the OPR for a pixel in The Brick ($l=0.268^\circ$, $b=0.028^\circ$, $T_{\mathrm{kin}}=78.9$~\K{}). 
    Red and blue contours show the $3\sigma$ (solid) and $4\sigma$ (dashed) confidence regions for o-H$_2$CO and p-H$_2$CO column densities as functions of H$_2$ density. 
    The vertical green line marks the adopted density ($10^5$~\cmc{}). 
    The dot-dashed lines indicate the column density ranges at this density ($3\sigma$ confidence). 
    The OPR (with uncertainty) is computed from the ratio of the best-fit column densities and is given in the bottom-left corner.
}
    \label{fig:opr}
\end{figure}

Figure~\ref{fig:opr} illustrates this methodology for a representative pixel. The confidence regions for o-H$_2$CO and p-H$_2$CO are shown in the $n_{\mathrm{H_2}}$--$N_{\mathrm{H_2CO}}$ plane. At the adopted density (vertical line), the respective best-fit column densities are obtained, and their ratio gives the OPR for that pixel.

We tested the effects of different prior OPR hypotheses ($\mathrm{OPR} = 1$, $2$, and $3$) on the results of three molecular cloud analyses. Statistical analysis revealed extremely low dependence of the derived OPR values on prior selection, with the OPR distributions of all three molecular clouds centered around the statistical equilibrium value of $3.0$.
For the representative case of prior $\mathrm{OPR} = 3$, the mean values are $3.26 \pm 1.43$ for The Brick, $3.09 \pm 1.20$ for Sgr~A1, and $2.80 \pm 1.04$ for Sgr~A2, where the uncertainties represent the standard deviation of the pixel-wise OPR 
distribution. 
This consistency validates our adoption of $\mathrm{OPR} = 3$ as a reasonable prior in the temperature and density analysis presented in Section~\ref{ssec:temp_results}. 
\citet{Tang2018} measured the H$_2$CO OPR in $\sim 100$ ATLASGAL-selected massive clumps in the Galactic disk, finding values ranging from $1.0$ to $3.0$ with a mean of $2.0 \pm 0.1$. In comparison, the higher OPR values ($\sim 3$) derived for our CMZ clouds suggest that the H$_2$CO molecules in these regions have reached thermal equilibrium, likely facilitated by the elevated temperatures and extreme physical conditions prevalent in the CMZ environment.

\section{Heating Mechanisms of Dense Gas in the CMZ Clouds}
\label{sec:discussion}
\subsection{Relationship Between Gas Temperature and Velocity Dispersion}
\label{subsec:temp_vdisp}
\begin{figure*}[ht!]
    \centering
    \includegraphics[width=0.85\textwidth]{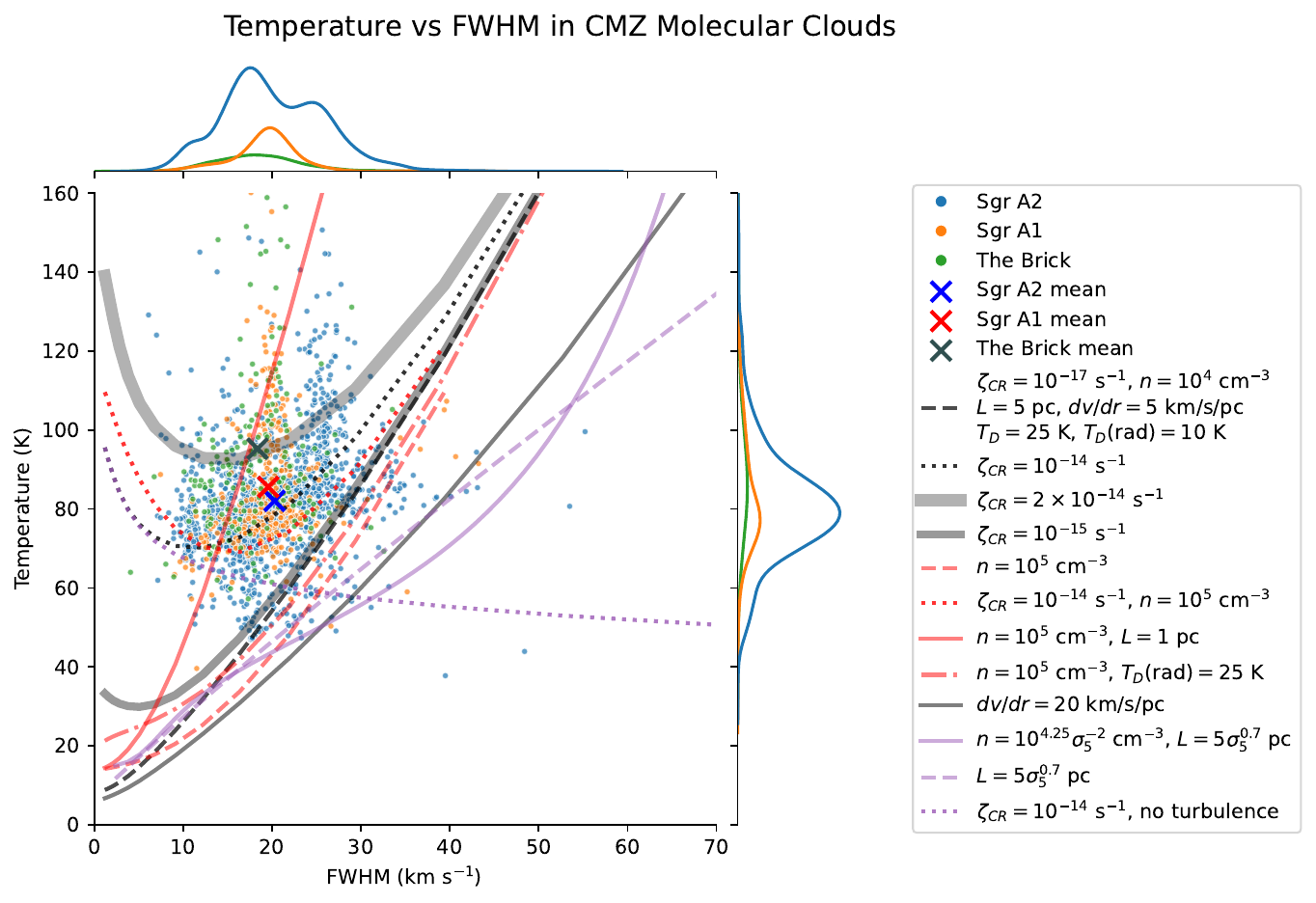}
    \caption{Relationship between the derived kinetic temperature ($T_{\mathrm{kin}}$) and the velocity dispersion (line width, FWHM) for the three target molecular clouds: Sgr~A2, Sgr~A1, and The Brick (G0.253+0.016). The mean kinetic temperature and velocity dispersion for each cloud are marked with ``$\times$'' symbols. Marginal distributions along the top and right axes display the kernel density estimates for the temperature and line width (FWHM) of each cloud, respectively, and employ a consistent color scheme matching the scattered points. The solid curves show model predictions from \textsc{despotic} \citep{Krumholz2014} under different physical assumptions.
    }
    \label{fig:TemperaturevsFWHM}
\end{figure*}

We first examine the relationship between $T_{\mathrm{kin}}$ and velocity dispersion. 
The thermal velocity dispersion for H$_2$CO at $T_{\mathrm{kin}} \sim 100$~\K{} is $< 1$~\kms{}, which is negligible compared to the observed line widths (FWHM $\sim 10$--$30$~\kms{}). Therefore, no thermal broadening correction was applied to the measured velocity dispersions. Following the approach of \citet{Ginsburg2016}, we configured multiple model scenarios and compared our observational data points against these models in Figure~\ref{fig:TemperaturevsFWHM}.

We use \textsc{despotic} \citep{Krumholz2014}\footnote{\url{https://despotic.readthedocs.io/en/latest/}} to compute the equilibrium gas temperature of molecular clouds under various physical conditions. \textsc{despotic} determines the steady-state gas temperature by solving the thermal balance equation, i.e., the temperature at which the total heating rate equals the total cooling rate. In our models, three main heating mechanisms are considered. The first is cosmic-ray heating: cosmic rays ionize H$_2$ molecules and deposit a fraction of their energy as heat into the surrounding gas; the heating rate is approximately proportional to the cosmic-ray ionization rate $\zeta_{\mathrm{CR}}$ and the gas number density $n$, so a higher $\zeta_{\mathrm{CR}}$ directly raises the heating rate. The second is turbulent dissipation heating: supersonic turbulence converts kinetic energy into thermal energy at the dissipation scale; the heating rate depends on the turbulent energy input rate, which in the \textsc{despotic} framework is jointly determined by the velocity gradient $\mathrm{d}v/\mathrm{d}r$, the velocity dispersion $\sigma_v$, and the cloud size $L$. A larger line width or a smaller cloud size implies a higher turbulent energy density per unit volume, thereby producing stronger heating. The third is dust--gas energy exchange: when the dust temperature differs from the gas temperature, the two exchange energy through collisions; in the CMZ, the dust temperature is generally lower than the gas temperature, so this process effectively acts as a cooling mechanism for the gas, and the effect of the radiation field on dust heating is parameterized by the radiation dust temperature $T_{\mathrm{D,rad}}$. The main cooling mechanisms include molecular line radiation and dust--gas collisional cooling. In our models, molecular line cooling is computed only from the rotational transitions of CO and its isotopologues ($^{13}$CO and C$^{18}$O), without including contributions from other species. This simplified treatment of molecular cooling differs slightly from the setup of \citet{Ginsburg2016} who additionally included atomic and molecular coolants such as O, C, C$^+$, H$_2$, and HD, but we have verified that this simplification has a negligible impact on the results: the overall trends of the model curves remain consistent, the temperature differences between the two sets of models are only a few kelvin within the linewidth range typical of our data (FWHM $\sim 10$--$30$~\kms{}), and reach $\lesssim$10~\K{} only at the smallest and largest line widths that lie beyond the bulk of our observations.

In Figure~\ref{fig:TemperaturevsFWHM}, we present the model-predicted temperature--line width relationships under different physical assumptions and compare them with the observational data. All model curves share a common set of baseline parameters: $\zeta_{\mathrm{CR}} = 10^{-17}$~\persec{}, $n = 10^4$~\cmc{}, $L = 5$~\pc{}, $\mathrm{d}v/\mathrm{d}r = 5$~\kmspc{}, $T_{\mathrm{D}} = 25$~\K{}, and $T_{\mathrm{D,rad}} = 10$~\K{}. Starting from this baseline, we vary one or more parameters at a time to investigate their effects on the equilibrium temperature. Increasing the cosmic-ray ionization rate ($\zeta_{\mathrm{CR}} = 10^{-15}$, $10^{-14}$, $2 \times 10^{-14}$~\persec{}) shifts the temperature curve upward as a whole, reflecting the role of cosmic rays as a pervasive heating source. Increasing the gas density to $n = 10^5$~\cmc{} enhances the efficiency of dust--gas collisional cooling. Decreasing the cloud size to $L = 1$~\pc{} causes turbulence to dissipate within a more compact region, raising the turbulent heating rate per unit volume. Increasing the radiation dust temperature to $T_{\mathrm{D,rad}} = 25$~\K{} reduces the cooling efficiency of dust--gas coupling, while increasing the velocity gradient to $\mathrm{d}v/\mathrm{d}r = 20$~\kmspc{} directly enhances the turbulent dissipation heating rate. We also adopt the parameterization introduced by \citet{Ginsburg2016} in which the density and cloud size scale with line width as $n = 10^{4.25}\,\sigma_5^{2}$~\cmc{} and $L = 5\,\sigma_5^{0.7}$~\pc{} (where $\sigma_5 = \sigma_v / 5$~\kms{}), corresponding to the assumption that clouds with larger line widths have higher virialized densities and larger spatial extents. Finally, we include a model curve without turbulent dissipation ($\zeta_{\mathrm{CR}} = 10^{-14}$~\persec{}, no turbulent heating) to demonstrate that cosmic-ray heating alone cannot reproduce the observed positive correlation between temperature and line width.

Our data show that the temperature distribution of the molecular clouds, particularly in the higher temperature regime, exhibits a significant discrepancy from model predictions that assume a low CRIR, such as the typical Galactic disk value of $\sim 10^{-17}$~\persec{} \citep[e.g.,][]{Farquhar1994}. In contrast, models with a higher CRIR (e.g., $\sim 10^{-14}$~\persec{}) provide a much better agreement with the observations. 
Specifically:

\begin{itemize}
    \item The $T_{\mathrm{kin}}$--FWHM distribution of the Sgr~A1 and Sgr~A2 regions aligns with the $\zeta = 1 \times 10^{-14}$~\persec{} model curves at number densities of $n = 10^4$ and $n = 10^5$~\cmc{}, a range that is consistent with the densities observed across all three molecular clouds in our study.
    \item Data points from The Brick (G0.253+0.016) are located nearer to the model curve for a smaller cloud scale ($L = 1$~\pc{}), or a higher CRIR ($\zeta = 2 \times 10^{-14}$~\persec{}). Additionally, we note that the highest-temperature regions in The Brick are spatially coincident with a likely shocked region in the southern part of the cloud \citep{Kauffmann2013, Johnston2014, Henshaw2019}.
\end{itemize}

 This comparison indicates that for these three clouds, pure turbulent dissipation may be insufficient to maintain the observed high-temperature state, suggesting the need for an additional heating source. Potential candidates include cosmic-ray heating, shock heating \citep[e.g.,][]{Johnston2014}, and localized heating by protostars \citep[e.g.,][]{Lu2017}. However, turbulent heating remains a crucial and likely non-negligible mechanism. The difficulty in matching the observed temperatures with high-CRIR models that exclude turbulent dissipation is clearly illustrated by the dotted purple curve in Figure~\ref{fig:TemperaturevsFWHM}, which shows the predicted temperature for $\zeta_{\mathrm{CR}} = 10^{-14}$~\persec{} without turbulent heating. Despite the elevated cosmic-ray ionization rate, this model predicts temperatures of only $\sim$50--70~\K{} at the line widths typical of our observations, significantly lower than the observed values. Furthermore, the curve remains essentially flat with respect to the velocity dispersion, failing to reproduce the observed positive correlation between $T_{\mathrm{kin}}$ and line width. This demonstrates that cosmic-ray heating alone, even at the high ionization rates inferred for the CMZ, cannot account for the observed temperatures or the temperature--line width relationship, and that turbulent dissipation is required as an additional heating mechanism. Its role accounts for a significant, and potentially dominant, fraction of the total heating budget. This remains true even when accounting for supplementary mechanisms such as cosmic-ray heating.

\subsection{Relationship Between Gas Temperature and Number Density}
\label{subsec:temp_density}
\begin{figure*}[ht!]
    \centering
    \includegraphics[width=0.85\textwidth]{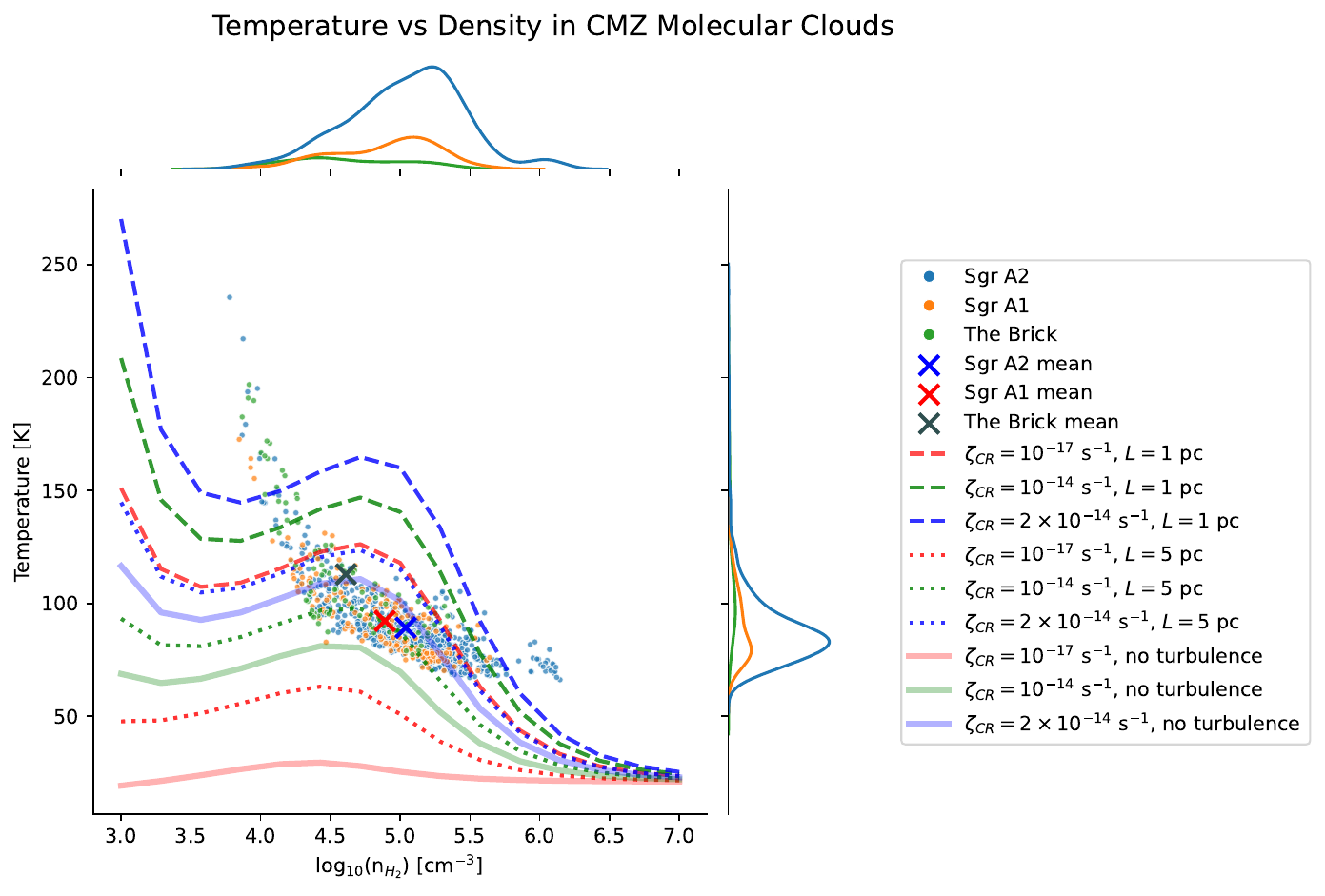}
    \caption{
        Relationship between the derived kinetic temperature ($T_{\mathrm{kin}}$) and the H$_2$ number density ($n_{\mathrm{H_2}}$) for pixels with well-constrained densities. Symbols are the same as in Figure~\ref{fig:TemperaturevsFWHM}. The curves show model predictions from \textsc{despotic}. Line color indicates the CRIR: $\zeta_{\mathrm{CR}} = 10^{-17}$~\persec{} (red), $10^{-14}$~\persec{} (green), and $2\times10^{-14}$~\persec{} (blue). Line style indicates the cloud scale and physics: $L=5$~\pc{} with turbulence (dotted), $L=1$~\pc{} with turbulence (solid), and a model without turbulent heating (dashed, $L=5$~\pc{}). Other parameters are held fixed ($\Delta V \approx 8.5$~\kms{}, $dv/dr = 5$~\kmspc{}, $T_D = 20$~\K{}).
    }
    \label{fig:TvsDensity}
\end{figure*}

To further elucidate potential differences in the heating mechanisms between Sgr~A1, Sgr~A2, and The Brick (G0.253+0.016), we analyze the relationship between the kinetic temperature ($T_{\mathrm{kin}}$) and the H$_2$ number density ($n_{\mathrm{H_2}}$), as shown in Figure~\ref{fig:TvsDensity}. The physical parameters for each pixel were derived from joint constraints in the density--column density--temperature parameter space using both the $J=3$--$2$ and $J=5$--$4$ spectral lines. The colored data points shown in the figure are selected from pixels identified as having well-constrained densities in Section~\ref{ssec:density_opr}; their corresponding $T_{\mathrm{kin}}$ and $n_{\mathrm{H_2}}$ values are calculated under the assumption of an H$_2$CO OPR of $3$. The subsequent analysis and comparison with models are based solely on this subset of reliable data, while results with poorly constrained densities are omitted from the figure and excluded from the discussion below.

For the model calculations, we fixed several parameters to characteristic values for these clouds: a linewidth of $\Delta V = 8.5$~\kms{}, a velocity gradient of $\mathrm{d}v/\mathrm{d}r = 5$~\kmspc{}, and a dust temperature of $T_{\mathrm{d}} = 20$~\K{}. We systematically varied two key parameters: the CRIR ($\zeta$) and the cloud scale ($L$). Additionally, we computed models both with and without the contribution of turbulent heating to isolate its effect.
In Figure~\ref{fig:TvsDensity}, the distribution of our reliable data points in the temperature--density plane shows a closer overall alignment with models that assume a high CRIR and include turbulent heating. This stands in clear contrast to models with a low cosmic-ray ionization rate or those without turbulent heating, a finding consistent with the conclusion drawn in the previous section. This agreement also demonstrates that our joint analysis, utilizing two distinct spectral line ratios, successfully provides well-constrained estimates for both density and column density in the higher-temperature regions.

\subsection{Comparison with Previous Studies and Implications for CMZ Heating}
\label{subsec:comparison}

Our study revises downward the previously reported extreme temperatures ($>100$~\K{}) in The Brick, Sgr~A1, and Sgr~A2. The resulting average kinetic temperatures of these clouds ($\sim 84$--$95$~\K{}, see Table~\ref{table:temperature_results}) remain systematically higher than the typical 60~\K{} found across the broader CMZ by \citet{Ginsburg2016}. This places them at the high end of, although not as extreme outliers from, the temperature distribution observed in the CMZ. Their sustained warmth suggests they are subject to enhanced heating, which likely involves mechanisms beyond the turbulent dissipation that is thought to be a primary and widespread heating source in CMZ clouds \citep{Ginsburg2016, Immer2016}.

The observed temperature elevation likely stems from a combination of mechanisms. First, localized mechanical processes are probable contributors. The localized temperature enhancements and the positive $T_{\mathrm{kin}}$ trend we observe (Figure~\ref{fig:TemperaturevsFWHM}) are consistent with turbulent dissipation as proposed by previous studies \citep{Pan2009, Ginsburg2016, Immer2016}. Furthermore, in clouds like The Brick, our temperature maps show that the highest-temperature regions coincide spatially with the shock-dominated areas identified by \citet{Johnston2014} and \citet{Henshaw2019}, supporting a local role for shock heating. Stellar feedback may provide additional heating in star-forming regions; our temperature estimates for Sgr~A1 and Sgr~A2 ($\sim 82$--$86$~\K{}) are broadly consistent with the localized protostellar heating suggested by \citet{Lu2017}, though lower than their earlier estimates ($\gtrsim 100$~\K{}) due to our improved multi-transition constraints.

In addition to these localized mechanisms, a pervasive, global heating source may be responsible for the systematically elevated temperature across our target clouds compared to the wider CMZ. Cosmic-ray heating is a leading candidate for such a background mechanism. The spatial location of our targets supports this: all three clouds reside near the Galactic center (within $|l| < 0.3^\circ$), a region with a high concentration of potential cosmic-ray sources, such as supernova remnants \citep{Heywood2022} and activity linked to Sgr~A* \citep{Marin2023}. This is consistent with a growing consensus that the CMZ possesses an elevated cosmic-ray ionization rate (CRIR) of $\sim 10^{-15}$--$10^{-14}$~\persec{} \citep[e.g.,][]{Oka2019, Indriolo2015, Rivilla2022, Sanz2024}, much higher than in the Galactic disk. Additionally, our own analysis finds that the observed temperatures are best matched by models incorporating both turbulent heating and a CRIR of $\sim 10^{-14}$~\persec{} (Section~\ref{subsec:temp_density}), linking the global heating to the cosmic-ray environment.

The concurrent role of cosmic-ray and mechanical heating finds a compelling analogy in the starburst galaxy NGC~253. Using multiple H$_2$CO transitions, \citet{Mangum2019} measured the gas temperature within giant molecular clouds in its central molecular zone. They derived $T_\mathrm{kin} \gtrsim 50$~\K{} on scales of $\sim 80$~\pc{}, with temperatures exceeding 300~\K{} on smaller scales ($\lesssim 16$~\pc{}). By analyzing the abundances of various molecules across these clouds, they found a chemical gradient consistent with a scenario dominated by cosmic-ray heating, yet one that also requires a significant contribution from mechanical (i.e., turbulent/shock) heating. This picture has been substantially reinforced by the ALCHEMI large program, which carried out spatially resolved multi-molecular diagnostics across NGC~253's CMZ. Using HCO$^+$/HOC$^+$ abundance ratios, \citet{Harada2021} inferred CRIRs of $\sim 10^{-14}$--$10^{-12}$~\persec{}, far exceeding the Galactic average. Analyses of C$_2$H \citep{Holdship2021} and H$_3$O$^+$/SO \citep{Holdship2022} independently confirmed that cosmic-ray heating dominates the energy budget in this environment. \citet{Behrens2022} further mapped the spatial variation of the CRIR using HCN/HNC ratios, finding that the ionization rate correlates positively with the concentration of supernova remnants, while \citet{Behrens2024} extended this analysis and confirmed a CRIR enhancement toward the central regions. This multi-diagnostic picture from NGC~253, where high temperatures are driven by a combination of pervasive cosmic-ray heating and localized mechanical processes, closely parallels the interpretation we propose for our Galactic Center clouds.

In summary, the elevated temperatures in our target clouds are best explained by the concurrent action of multiple heating processes. A pervasive background of cosmic-ray heating, enhanced in the central CMZ, may raise the overall thermal floor. Superposed on this are strong, localized contributions from turbulent dissipation and shock heating, which can account for internal temperature structures and hotspots, while stellar feedback may provide further local input in star-forming regions. This multi-mechanism framework consistently accounts for both the overall warm state and the complex internal temperature patterns of the dense gas in the CMZ.

\section{Conclusions}
\label{sec:conclusions}
In this study, we conducted a joint analysis of the H$_2$CO $J=3$--$2$ and $J=5$--$4$ transitions, combined with non-LTE radiative transfer modeling, to constrain the kinetic temperature of three prominent molecular clouds (The Brick, Sgr~A1, and Sgr~A2) in the CMZ and to investigate their heating mechanisms. Our principal findings are summarized as follows:
\begin{enumerate}
    \item The joint use of H$_2$CO $J=3$--$2$ and $J=5$--$4$ lines significantly improves the accuracy of temperature estimates in warm molecular clouds. Using the $J=3$--$2$ data alone leads to a systematic overestimation of temperature in high-temperature regimes ($T_{\mathrm{kin}} > 100$~\K{}) due to level thermalization, with average temperatures reaching $124.6$~\K{}, $111.4$~\K{}, and $95.9$~\K{} for The Brick, Sgr~A1, and Sgr~A2, respectively. Conversely, the $J=5$--$4$ data alone often overestimates temperatures in low-density regions. Our joint analysis effectively overcomes the limitations of individual datasets, yielding more reliable kinetic temperatures of $94.0$~\K{}, $85.6$~\K{}, and $82.1$~\K{} for The Brick, Sgr~A1, and Sgr~A2 (assuming an ortho-to-para ratio (OPR) of $3$), respectively. 
    \item Our joint analysis demonstrates robustness against the uncertain OPR of H$_2$CO. The derived kinetic temperatures show negligible sensitivity to the assumed OPR value within the range of $1.0$--$3.0$, with variations significantly smaller than uncertainties from other model parameters. We constrained the OPR from our data and find mean values of $3.26 \pm 1.43$, $3.09 \pm 1.20$, and $2.80 \pm 1.04$ for The Brick, Sgr~A1, and Sgr~A2, respectively. These values are consistent with the statistical equilibrium value of $3$, confirming that H$_2$CO in these clouds is in thermal equilibrium and validating our adoption of OPR $= 3$ for the main analysis.
    \item Analysis of the relationship between temperature, velocity dispersion, and number density suggests that the three molecular clouds are heated by multiple heating mechanisms. The observational data are consistent with models that invoke a high CRIR ($\zeta \sim 1$--$2 \times 10^{-14}$~\persec{}), and exhibit significant discrepancies with predictions from pure turbulent heating models. 
\end{enumerate}

Our results demonstrate that joint multi-transition line analysis is essential for accurate gas temperature determination in extreme environments like the CMZ. This work provides new observational constraints for understanding the energy balance in the CMZ. Future observations with higher resolution and sensitivity (e.g., with ALMA, JWST) will be crucial to further disentangle the relative contributions of cosmic-ray and turbulent heating.

\section*{Acknowledgments}
Y.A. acknowledges the support from the National Natural Science Foundation of China (NSFC) (Grant No. 12173089), and the China Manned Space Program with grant no. CMS-CSST-2025-A10. X.L. acknowledges the support from the National Key R\&D Program of China (No.\ 2022YFA1603101), NSFC through grant Nos.\ 12273090 and 12322305, the Natural Science Foundation of Shanghai (No.\ 23ZR1482100). Y.A. and X.L. acknowledge the support from the Strategic Priority Research Program of the CAS Grant No.\ XDB0800300 and the CAS ``Light of West China'' Program No.\ xbzg-zdsys-202212.\\
These observations were obtained by the James Clerk Maxwell Telescope, operated by the East Asian Observatory on behalf of The National Astronomical Observatory of Japan; Academia Sinica Institute of Astronomy and Astrophysics; the Korea Astronomy and Space Science Institute; the National Astronomical Research Institute of Thailand; Center for Astronomical Mega-Science (as well as the National Key R\&D Program of China with No. 2017YFA0402700). Additional funding support is provided by the Science and Technology Facilities Council of the United Kingdom and participating universities and organizations in the United Kingdom and Canada.\\

\software{
    Astropy \citep{astropy2022},
    Despotic \citep{Krumholz2014},
    matplotlib \citep{Hunter2007},
    myRadex \citep{myRadex},
    numpy \citep{numpy2020},
    ORAC-DR \citep{Jenness2015},
    pandas \citep{pandas2020},
    pyspeckit \citep{Ginsburg2011,Ginsburg2022},
    RADEX \citep{Van2007},
    scipy \citep{scipy2020},
    Starlink \citep{Currie2014}
}

\appendix

\section{Baseline Fitting}
\label{app:baseline}
\renewcommand{\thefigure}{A\arabic{figure}}
\setcounter{figure}{0}

\begin{figure*}[ht!]
    \centering
    \includegraphics[width=\textwidth]{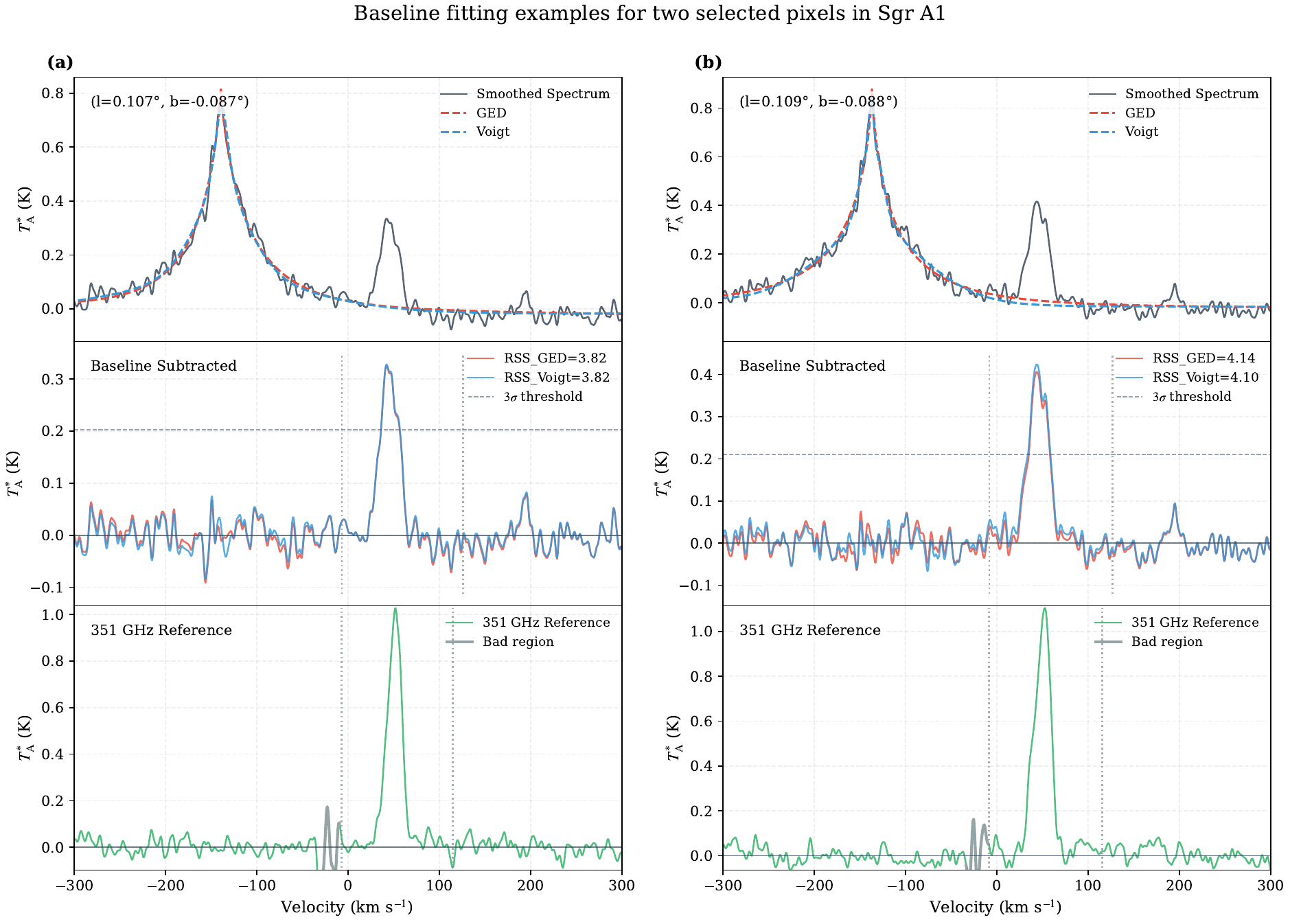}
    \caption{
    Representative examples of the instrumental baseline fitting and subtraction process for the H$_2$CO $5_{33}$--$4_{32}$ ($364$~\GHz{}) line data.
    (a) Fit for the pixel at ($l=0.107^\circ$, $b=-0.087^\circ$) in cloud Sgr~A1. Both the  generalized exponential distribution function (GED) and Pseudo-Voigt models provide excellent fits to the instrumental baseline (residual sum of squares, $\mathrm{RSS} \approx 3.82$). The middle panel shows the successfully subtracted spectrum, and the bottom panel shows the corresponding $351$~\GHz{} spectrum used to define the masked velocity range.
    (b) Baseline fitting for a nearby pixel at ($l=0.109^\circ$, $b=-0.088^\circ$) in Sgr~A1, demonstrating the consistency of the baseline structure. The Pseudo-Voigt model provides a marginally better fit ($\mathrm{RSS} = 4.10$) than the GED model ($\mathrm{RSS} = 4.14$) here.
    Top panels: The observed spectrum (black) and the two fitted baseline models (GED in red, Pseudo-Voigt in blue).
    Middle panels: The final baseline-subtracted spectrum. The horizontal dashed lines indicate the $y=0$ and $y=\pm3\sigma$ levels.
    Bottom panels: The corresponding $351$~\GHz{} reference spectrum, used to identify and mask channels (the region between two vertical lines) containing potential astrophysical emission, which are excluded from the baseline fit.
    The RSS values for each fit are provided.
    }
\label{fig:baseline_examples}
\end{figure*}

The $364$~\GHz{} lines are significantly weaker than the $351$~\GHz{} line and sit atop an anomalous, non-flat baseline caused by instrumental effects (e.g., mixer instability) or atmospheric effects; see Figure~\ref{fig:baseline_examples}. Precise modeling and subtraction of this baseline are essential for accurate measurement of the $364$~\GHz{} lines' astrophysical parameters (intensity, width, velocity). In contrast, the $351$~\GHz{} line, having a higher signal-to-noise ratio, exhibits a region of anomalously increased noise near the line profile, which must be excluded from analysis.

This appendix details the full procedure for subtracting the instrumental baseline from the H$_2$CO $5_{33}$--$4_{32}$ and $5_{32}$--$4_{31}$ ($\sim 364$~\GHz{}) spectral line data cube.

1. $351$~\GHz{} Line Preprocessing: The $351$~\GHz{} line data were first slightly smoothed with a Gaussian kernel ($\sigma = 2$ channels) to improve signal-to-noise ratios. The noise root mean square ($\mathrm{RMS}_{351}$) was then calculated by examining channels far from line emission.

2. Bad Channel Masking and Emission Interval Definition: Based on the smoothed $351$~\GHz{} line, channels with intensities exceeding $5\times$ $\mathrm{RMS}_{351}$ were preliminarily identified as potential emission. Considering the large velocity dispersion typical of sources in our observed CMZ region (typical FWHM $\sim 25$~\kms{}), this preliminary identification region was conservatively expanded: the maximum and minimum supra-threshold velocity points were extended by approximately 50 channels ($\sim 40$~\kms{}) on either side to ensure complete coverage of all possible emission components. Known bad channels in the $351$~\GHz{} line were manually flagged as NaN and excluded from subsequent analysis.

3. Velocity Alignment: Emission Interval Mapping to the $364$~\GHz{} Line: The emission velocity interval defined above was mapped onto the velocity axis of the $364$~\GHz{} data. Because the two blended $364$~\GHz{} transitions ($5_{33}$--$4_{32}$ and $5_{32}$--$4_{31}$) have slightly different rest frequencies, their velocity coordinates are offset from each other by $\Delta v$. To ensure that the masked interval covers emission from both transitions, the lower velocity bound was set by the higher-frequency transition, and the upper velocity bound was extended by $\Delta v$ to account for the lower-frequency transition. The resulting interval ($V_{\mathrm{mask, 364}}$) was excluded from subsequent baseline fitting.

4. Baseline Modeling and Fitting:

For the $351$~\GHz{} line, its data were divided into two parts:

(a) Fitting Interval: Channels whose velocity range lies outside $V_{\mathrm{mask, 364}}$, deemed to contain only baseline signal and noise.

(b) Masked Interval: Channels whose velocity range lies inside $V_{\mathrm{mask, 364}}$, potentially containing astrophysical emission, not used during baseline fitting.

We tested various functional forms to model the baseline and ultimately found that a pseudo-Voigt function combined with a generalized exponential distribution function (GED) performed best overall. For the majority of pixels, the baseline subtraction results from both modeling approaches are nearly identical, noticeable differences occur only in a small subset of pixels (Figure~\ref{fig:baseline_examples}). Based on visual inspection, we find that the pseudo-Voigt model yielded marginally better fits in regions with complex baseline structure than the GED model, and was therefore selected as the final method for baseline removal across the entire data cube. The complete model was then subtracted from the $364$~\GHz{} line data, yielding the baseline-subtracted net spectrum.

The mathematical expression of the pseudo-Voigt function is as follows:

\begin{equation}
f(x) = A_1 \left[ \mu_1 L_1(x) + (1 - \mu_1) G_1(x) \right] + A_2 \left[ \mu_2 L_2(x) + (1 - \mu_2) G_2(x) \right] + P(x)
\end{equation}

where:

\begin{equation}
L_i(x) = \frac{2}{\pi} \cdot \frac{w_{Li}}{4(x - x_{c1})^2 + w_{Li}^2} 
\end{equation}
\begin{equation}
G_i(x) = \frac{\sqrt{4 \ln 2}}{\sqrt{\pi} w_{Gi}} \cdot \exp\left( -\frac{4 \ln 2}{w_{Gi}^2} (x - x_{c1})^2 \right) 
\end{equation}
\begin{equation}
P(x) = y_0 + a x^2 + b x
\end{equation}

Parameter definitions:
\begin{itemize}
\item $x$: Independent variable (frequency or velocity)
\item $x_{c1}$: Central position of the spectral line (shared by both pseudo-Voigt functions)
\item $A_1, A_2$: Amplitudes of the two pseudo-Voigt functions
\item $\mu_1, \mu_2 \in [0, 1]$: Mixing parameters controlling the proportion of Gaussian and Lorentzian components ($\mu_i = 1$ represents a pure Lorentzian function, $\mu_i = 0$ represents a pure Gaussian function)
\item $w_{L1}, w_{L2}$: Full width at half maximum (FWHM) of the Lorentzian components
\item $w_{G1}, w_{G2}$: Full width at half maximum (FWHM) of the Gaussian components
\item $y_0, a, b$: Coefficients of the quadratic background polynomial
\end{itemize}

\section{Example Spectra for Representative Pixels}
\label{app:fits}
\renewcommand{\thefigure}{B\arabic{figure}}
\setcounter{figure}{0}

In this appendix, we present spectral line plots for three representative molecular cloud regions (The Brick, Sgr~A1, and Sgr~A2) across four different molecular transitions. These spectral lines include the H$_2$CO transitions: 3$_{03}$--2$_{02}$ (218.222190~\GHz{}), 3$_{21}$--2$_{20}$ (218.760070~\GHz{}), 5$_{15}$--4$_{14}$ (351.768645~\GHz{}), and the blended 5$_{33}$--4$_{32}$ and 5$_{32}$--4$_{31}$ lines (364.275141~\GHz{} and 364.288884~\GHz{}). For each region, a representative spatial position was selected for analysis. The $J=3$--$2$ transition data were obtained from \citet{Ginsburg2016}, and the spectra were truncated to a velocity range of $300$~\kms{}. The $J=5$--$4$ transition data were extracted over the same velocity range for direct comparison. The vertical axis shows the antenna temperature $T_{\mathrm{A}}^{*}$ (~\K{}). The observed spectra are shown as black solid lines, with spectral line identifications marked at appropriate positions. 

\begin{figure}[ht!]
    \centering
    \includegraphics[width=\textwidth]{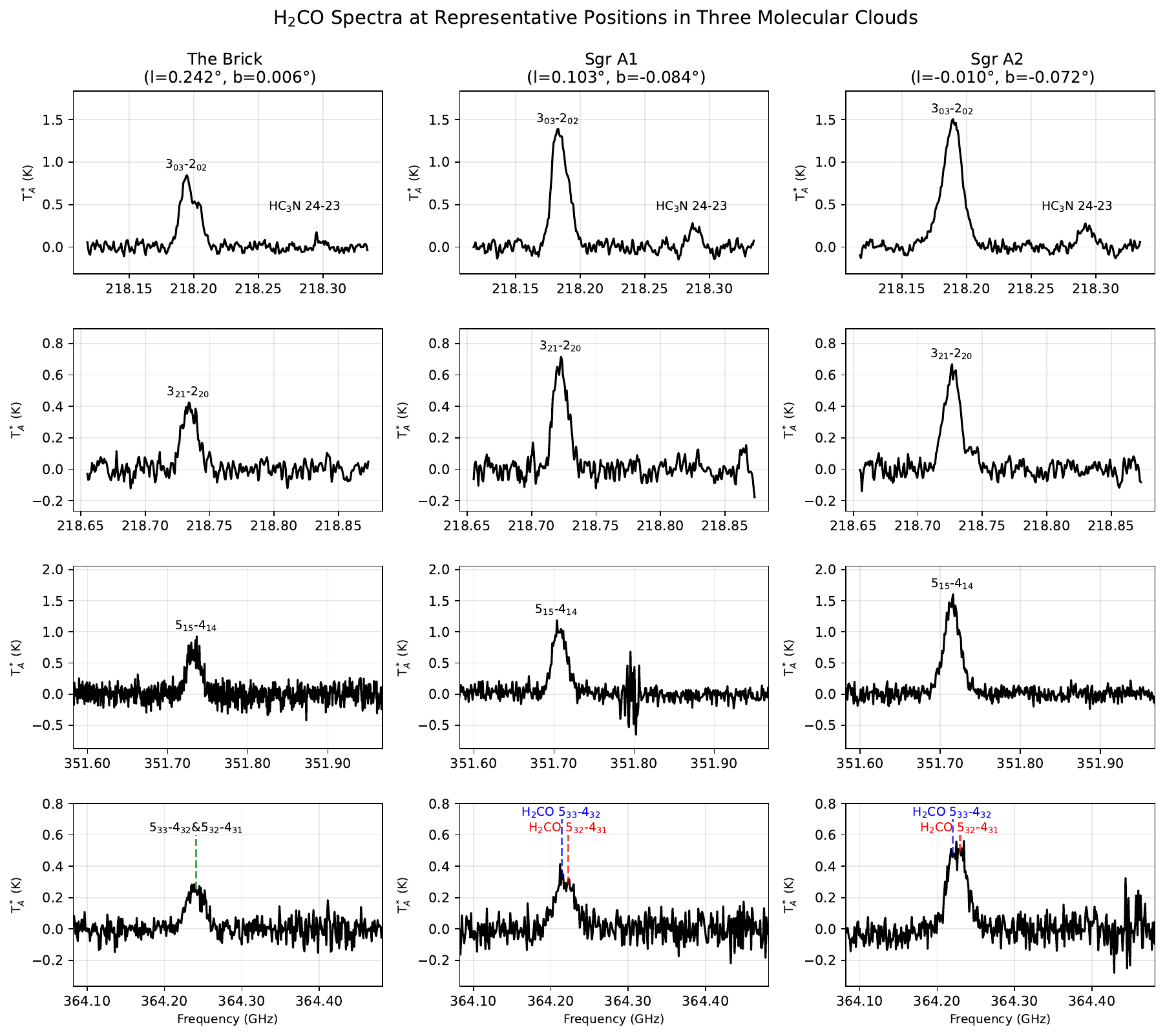}
    \caption{
        Spectral line plots of four H$_2$CO molecular transitions for three representative molecular cloud regions (The Brick, Sgr~A1, and Sgr~A2).
        Row 1: H$_2$CO 3$_{03}$--2$_{02}$ line ($\nu_0 = 218.222190$~\GHz{});
        Row 2: H$_2$CO 3$_{21}$--2$_{20}$ line ($\nu_0 = 218.760070$~\GHz{});
        Row 3: H$_2$CO 5$_{15}$--4$_{14}$ line ($\nu_0 = 351.768645$~\GHz{});
        Row 4: Blended lines of H$_2$CO 5$_{33}$--4$_{32}$ and 5$_{32}$--4$_{31}$ ($\nu_0 = 364.275141$~\GHz{} and $364.288884$~\GHz{}).
        The horizontal axis of each panel shows frequency (\GHz{}), while the vertical axis shows antenna temperature $T_{\mathrm{A}}^{*}$ (~\K{}). Major spectral lines are identified in the plots. 
        For the H$_2$CO 3$_{03}$--2$_{02}$ line, we also mark the position of the HC$_3$N 24--23 line.
        Note that the H$_2$CO 5$_{33}$--4$_{32}$ and 5$_{32}$--4$_{31}$ lines are closely spaced, and for the selected Brick spectrum, we do not separate these two lines.
    }
    \label{fig:A1_spectra}
\end{figure}

\clearpage

\bibliography{reference}{}
\bibliographystyle{aasjournalv7}



\end{document}